\documentclass[AEJ]{AEA}

\input{tcilatex}
\draftSpacing{1.5}

\usepackage{array}
\usepackage{natbib}
\usepackage{graphicx} 
\usepackage{multirow} 
\usepackage{textcomp}
\usepackage{gensymb}
\usepackage[normalem]{ulem}
\begin{document}

\date{%
\today%
}
\title{The Effects of Climate and Weather on Economic Output: Evidence from Global Subnational Data}
\shortTitle{Dong ET AL: The effects of climate conditions}
\author{Jinchi Dong, Richard S.J. Tol and Jinnan Wang\thanks{Dong: Ma Yinchu School of Economics, Tianjin University, Tianjin, China;
(dongjinchi@163.com). 
Tol: Department of Economics, University of Sussex, Falmer, United Kingdom;
Institute for Environmental Studies, Vrije Universiteit, Amsterdam, The Netherlands; 
Department of Spatial Economics, Vrije Universiteit, Amsterdam, The Netherlands;
Tinbergen Institute, Amsterdam, The Netherlands;
CESifo, Munich, Germany;
Payne Institute for Public Policy, Colorado School of Mines, Golden, CO, USA;
(R.Tol@sussex.ac.uk).
Wang: Chinese Academy of Environmental Planning, Beijing, 100043, China;
(wangjn@caep.org.cn). We thank Professor Shqiponja Telhaj for her helpful comments and discussions.}}

\begin{abstract}
Estimating the effects of climate on economic output is crucial for formulating climate policy, but current empirical findings remain ambiguous. Using annual panel model and panel long-difference model with global subnational data from nearly all countries, we find robust evidence that weather shocks have a \emph{transient} effect on output. The impact on economic growth is large and significant in the short-run but statistically insignificant in the long-run, except in the coldest and hottest places.   
\end{abstract}

\maketitle

Understanding the effects of climate conditions on economic output is crucial for analyzing the social cost of carbon \citep{tol2011social}, the inequitable impact of climate change \citep{anthoff2019inequality, Chancel2025}, and the physical risks to financial institutions \citep{jongman2014increasing}. They also play a central role in quantifying potential compensation for ``loss and damage" \citep[e.g., ][]{tol2004, burke2023quantifying}. Although extensive research has been conducted on the economic impact of weather and climate, effect size and mechanism are disputed \citep{Dell2014, Kolstad2020, Polonik2025}.

A key point of debate in these studies centers on the so-called ``level vs growth" problem. If weather affects its level, economic output declines during anomalous years but rebounds once the weather returns to prevailing conditions. In contrast, if weather affects growth, the decline in output persists after conditions revert to the prevailing norm. If climate change directly affects economic growth, the projected future economic output losses are over twenty times greater than if the impact is on levels \citep{tol2024meta} and the social cost of carbon is \$265/tC higher \citep{Tol2025anyas}.

Three things complicate the level vs growth debate. First, a drop in output manifests itself as a drop in growth. Second, a drop in output affects investment and hence growth \citep{Nordhaus1993, fankhauser2005climate}. Third, the "return to prevailing conditions" is a return to the climate of 1960 in some papers \citep{dell2012temperature, burke2015global, pretis2018, Henseler2019, Acevedo2020, damania2020does, kalkuhl2020impact, Kikstra2021, Callahan2025, Neal2025}, a return to the climate of the past 30 years in other papers \citep{kahn2021long, Ahmadi2025, Apergis2025, Tolfc}, or a return to the climate extrapolated from the recent past in yet other papers \citep{Bilal2024, Muller2025}. The difference is subtle in sample, but large out of sample \citep{newell2021gdp} and the interpretation of coefficients is very different \citep{merel2021climate}.

Year-to-year variations capture the short-term impact of the weather on the economy rather than the long-term effects of the climate. Adaptation is the main difference \citep{burke2016adaptation, auffhammer2018quantifying, hsiang2016climate}. Complete adaptation implies a transient (level) effect, incomplete adaptation a permanent (growth) effect. Weather data are more plentiful than climate data and arguably orthogonal to economic data. Suggested solutions include long differences \citep{burke2016adaptation}, comparison to cross-sections \citep{dell2009}, spatial models \citep{Deryugina2017}, simultaneous models \citep{Tolfc}, and multilevel models \citep{Auffhammer2022}. We here use long differences.

Most studies rely on national data and use a limited number of control variables, relying on country fixed effects instead. Estimates may be biased if uncontrolled national characteristics are correlated with climate. Subnational data counter this and are used by \citet{damania2020does, kalkuhl2020impact, kotz2021day, kotz2022effect} and \citet{kotz2024} to study the impact of weather shocks on economic growth. Unfortunately, the data used by these studies exclude many hot and poor countries. The average temperature and GDP per capita for the 77 countries in \citet{kalkuhl2020impact} are 10.4\celsius{} and \$12,334, while the global averages are around 13.6\celsius{} and \$11,194. Since poor and hot regions are expected to be more vulnerable to weather and climate, using incomplete data may underestimate the effects of weather and climate. Only \citet{damania2020does} have global coverage \footnote{ \citet{meierrieks2024temperature} also have global coverage, but they studies the impact of climate on economic income rather than output.}.

Heterogeneity is a second controversy in the literature: Are only poor regions negatively affected \citep{dell2012temperature}, all countries \citep{kahn2021long}, or hot countries negatively and cold countries positively \citep{burke2015global}?

In this paper, we return to the level vs growth debate using subnational data. Unlike the studies by Kalkuhl and Kotz, we have global coverage: We use GDP data for over 1600 regions in 196 countries. Unlike the paper by Damania, we \emph{test} level specifications against growth models and we consider heterogeneity between countries. Furthermore, we show that the missing observations in the Kalkuhl database affect their results.

The annual panel model reveals a significant effect of temperature on output growth, but no effect on output. The optimal temperature implied by the model is 14.6\celsius. The effect of temperature is nearly identical in rich and poor regions. A 1\celsius{} temperature increase at 25\celsius{} decreases GDP per capita growth by 1.7\% in poor regions and 1.5\% in rich regions. Poor countries show a larger response mainly because they are hotter on average, not because they are poorer. We also find a positive marginal effect of precipitation on output growth. 100mm increase in precipitation increases GDP per capita growth by around 0.1\%. Overall, our results support the growth effects of weather conditions on economic output rather than level effects.

The long-difference model shows insignificant effects of precipitation on output. In contrast, temperature has significantly negative effects on output growth, but only in extremely hot regions. A 1\celsius{} increase in temperature decreases decadal GDP per capita growth by 33.1\% at 29\celsius{}. This effect is the same for rich and poor regions.

GDP per capita in 2100 is projected to decrease 34.5\% (weighted by population) to 35.6\% (weighted by baseline GDP per capita) for 2.0\celsius{} warming based on the annual panel estimates. The 90th percentiles are -5.7\% and -3.5\%. Based on the long-difference estimates, the change in global average GDP per capita is statistically insignificant. Looking only at weather shocks overstates the impact of climate change. 

The remainder of the paper is organized as follows: Section I introduces the annual panel model and long-difference model used in this study. Section II describes the empirical approach. Section III introduces the data and provides descriptive statistics. Section IV presents our main results based on both the annual panel and long-difference models. Section V projects the future output loss based on our panel and long-difference estimates. Section VI provides a series of robustness checks. Section VII concludes the paper.

\section{Economic Model}
Following \citet{dell2012temperature}, we consider a production function that includes the effect of weather conditions on output per capita:
\begin{equation}
\label{eq:level}
y_{i t}=e^{c_{i}+\alpha_{0} T_{i t}^{2}+\beta_{0} T_{i t}}A_{i t}
\end{equation}
where $y_{i t}$ is the GDP per capita in the region $i$ and year $t$. $c_{i}$ captures regional fixed factors that affect the level of output. Following current empirical findings \citep{burke2015global, kalkuhl2020impact}, we consider the non-linear effects of weather conditions $T_{i t}$ on GDP per capita\footnote{We also employ a non-parametric method to check the nonlinear relationship between weather conditions and output growth.}. $A_{i t}$ measures total factor productivity.

Weather may also affect the growth rate of total factor productivity \citep{fankhauser2005climate, letta2019weather}. Therefore, we have:
\begin{equation}
\label{eq:growth}
\Delta ln(A_{i t}) = g_{i} + \gamma_{0} T_{i t}^{2}+\delta_{0} T_{i t}
\end{equation}
where $g_{i}$ captures regional fixed factors that affect productivity growth.

Taking the logarithm of Equation (\ref{eq:level}) and differencing with respect to time, we derive the growth equation:
\begin{equation}
\label{eq:firstdiff}
g_{i t} = ln(y_{i t})-ln(y_{i t-1})
= \alpha_{0} \Delta T_{i t} T_{i t}+ \alpha_{0} \Delta T_{i t} T_{i t-1} + \beta_{0}\Delta T_{i t} +\Delta  ln({A_{i t}})
\end{equation}

Substituting (\ref{eq:growth}) into (\ref{eq:firstdiff}) yields:
\begin{equation}
\label{eq:empiricalpanel}
g_{i t} = g_{i} +\alpha_{0} \Delta T_{i t} T_{i t}+ \alpha_{0} \Delta T_{i t} T_{i t-1}+ \beta_{0}\Delta T_{i t}+\gamma_{0} T^{2}_{i t}+\delta_{0} T_{i t}
\end{equation}
Equation (\ref{eq:empiricalpanel}) is the panel model proposed by \citet{kalkuhl2020impact} and used here to separately identify the level and growth effects of \textit{weather} conditions on output. Kalkuhl's estimates are based on subnational data from 77 countries. We re-estimate this equation using subnational data from 196 countries. The ``level effects" of weather conditions on output, which come from Equation (\ref{eq:level}), appear through $\alpha_{0}$ and $\beta_{0}$. The ``growth effects" of weather conditions, which come from Equation (\ref{eq:growth}), appear through $\gamma_{0}$ and $\delta_{0}$.

For the derivation of the long-difference model, we first take the average of outcomes and weather conditions over $m$ years in Equation (\ref{eq:level}) and Equation (\ref{eq:growth}) to capture the impact of climate. The relationship between average output per capita and climate, as well as the average total factor productivity and climate are then given by:
\begin{equation}
\label{eq:meanlevel}
ln (\overline{y_{i p}}) = c_{i}+\alpha_{0} \overline{T_{i p}}^{2}+\beta_{0} \overline{T_{i p}}+ln(\overline{A_{i p}})
\end{equation}
\begin{equation}
\label{eq:meangrowth}
\Delta ln(\overline{A_{i p}}) = g_{i} + \gamma_{0} \overline{T_{i p}}^{2}+\delta_{0} \overline{T_{i p}}
\end{equation}

For a specific period $p$ with $m$ years, Equation (\ref{eq:meanlevel}) serves as the cross-section model for assessing climate impacts. If uncontrolled, $c_{i}$ would bias results. To eliminate $c_{i}$, we take the difference of Equation (\ref{eq:meanlevel}) over two $m$-year periods:
\begin{equation}
\label{eq:longdiff}
\begin{aligned}
\overline{g_{i p_{n}}}&=ln (\overline{y_{i p}})-ln (\overline{y_{i p-n}}) \\ &= (c_{i}-c_{i})+\alpha_{0} (\overline{T_{i p}}^{2}- \overline{T_{i p-n}}^{2})+\beta_{0} (\overline{T_{i p}}-\overline{{T_{i p-n}}})+(ln(\overline{A_{i p}})- ln(\overline{A_{i p-n}})) \\ & = \alpha_{0} \Delta \overline{T_{i p_{n}}} \cdot \overline{T_{i p}}+\alpha_{0} \Delta \overline{T_{i p_{n}}} \cdot \overline{T_{i p-n}}+\beta_{0} \Delta \overline{T_{i p_{n}}} + \Delta ln(\overline{A_{i p_{n} }})
\end{aligned}
\end{equation}
We consider $n$ period differences to capture the long-term effect of climate, that is, $n$ is the gap between two periods.${g_{i p_{n}}}$ is the interperiod output growth. $\Delta \overline{T_{i p_{n}}}$ is interperiod climate change. $\Delta ln(\overline{A_{i p_{n} }})$ is interperiod total factor productivity growth.

According to Equation (\ref{eq:meangrowth}), the relationship between interperiod total factor productivity and climate is given by:
\begin{equation}
\label{eq:diffgrowth}
\begin{aligned}
\Delta ln(\overline{A_{i p_{n} }})&=\sum_{j=0}^{n-1} \Delta ln(\overline{A_{i p-j}}) \\
&=ng_{i}+\gamma_{0} \overline{T_{i p}}^{2}+\cdots+\gamma_{0} \overline{T_{i p-n+1}}^{2}+\delta_{0} \overline{T_{i p}}+\cdots+\delta_{0}\overline{T_{i p-n+1}}
\end{aligned}
\end{equation}

Substituting Equation (\ref{eq:diffgrowth}) into Equation (\ref{eq:longdiff}) yields:
\begin{equation}
\label{eq:longdiff1}
\begin{aligned}
\overline{g_{i p_{n}}}=& \alpha_{0} \Delta \overline{T_{i p_{n}}}^{2}+\beta_{0} \Delta \overline{T_{i p_{n}}} + \Delta ln(\overline{A_{i p_{n} }}) \\ = &ng_{i}+\alpha_{0} \Delta \overline{T_{i p_{n}}} \cdot \overline{T_{i p}}+\alpha_{0} \Delta \overline{T_{i p_{n}}} \cdot \overline{T_{i p-n}}+\beta_{0} \Delta \overline{T_{i p_{n}}}+ \\ & \gamma_{0} \overline{T_{i p}}^{2}+\cdots+\gamma_{0} \overline{T_{i p-n+1}}^{2}+\delta_{0} \overline{T_{i p}}+\cdots+\delta_{0}\overline{T_{i p-n+1}}
\end{aligned}
\end{equation}
Equation (\ref{eq:longdiff1}) is a generalization of the long-difference model proposed by \citet{burke2016adaptation}, where $\alpha_{0}$ and $\beta_{0}$ capture the ``level effects" of climate conditions on output and $\gamma_{0}$ and $\delta_{0}$ capture the ``growth effects" of climate conditions. This is the long-difference model used here to identify the level and growth effects of \textit{climate} conditions on output separately.

To achieve the ``long difference" over two periods, we can increase $n$ (the gap) or $m$ (the years of a period). For example, to obtain a 6-year difference, we can set $n=1$ and $m=6$ or $n=2$ and $m=3$. \footnote{Considering the years from 1990 to 2001, if $m=6$, periods would be 1990-1995 and 1996-2001. The difference between these two periods is 6 years when $n=1$. Alternatively, if $m=3$ and $n=2$, periods would be 1990-1992, 1993-1995, 1996-1998. The difference would be between 1990-1992 and 1996-1998, also resulting in a 6-year gap.} However, increasing $n$ leads to more lags in Equation (\ref{eq:longdiff1}). We set $n=1$. In this case, Equation (\ref{eq:longdiff1}) simplifies to:
\begin{equation}
\label{eq:emplongdiff}
\begin{aligned}
\overline{g_{i p}}=& \alpha_{0} \Delta \overline{T_{i p}}^{2}+\beta_{0} \Delta \overline{T_{i p}} + \Delta ln(\overline{A_{i p }}) \\ = &g_{i}+\alpha_{0} \Delta \overline{T_{i p}} \cdot \overline{T_{i p}}+\alpha_{0} \Delta \overline{T_{i p}} \cdot \overline{T_{i p-1}}+\beta_{0} \Delta \overline{T_{i p}}+ \gamma_{0} \overline{T_{i p}}^{2}+\delta_{0} \overline{T_{i p}}
\end{aligned}
\end{equation}
Where $\overline{g_{i p}}$ is the output growth between period $p$ and period $p-1$. $\Delta \overline{T_{i p}}$ is the change in climate conditions between the adjacent periods. Equation (\ref{eq:emplongdiff}) is the long-difference model used in this study.

\section{Empirical Model}
\subsection{Annual Panel Model}
We first employ an annual panel model to estimate the short-term effects of weather conditions on output, based on Equation (\ref{eq:empiricalpanel}). Like \citet{kalkuhl2020impact}, we ignore the lag term of $\Delta T_{i t} T_{i t-1}$ to provide a parsimonious regression model. We add it back in the robustness tests (Table \ref{tbl:altlag}). The regression specification is:
\begin{equation}
\label{reg: panelemp}
g_{i t} = \alpha\mathbf{\Delta T_{i t} T_{i t}}+ \beta\mathbf{\Delta T_{i t}}+\gamma \mathbf{T^{2}_{i t}}+\delta\mathbf{T_{i t}}+X_{it}+\eta_{i}+\theta_{t}+h_{i}(t)+\epsilon_{i t}
\end{equation}
where $g_{i t}$ is the annual GDP per capita growth in region $i$ and year $t$. $\mathbf{T_{it}}$ = ($T_{i t}, P_{i, t}$) is a vector of annual mean temperature (in \celsius) and annual total precipitation (in m). $X_{it}$ are control variables, including regional urbanization rate, mean years of schooling, and population, which might also influence output growth. $\eta_{i}$ is the region fixed effects, which controls for any unobserved, time-invariant differences between regions such as differing mean climate regimes and different baseline growth rates owing to geopolitical and historical factors. $\theta_{t}$ are the year fixed-effects, which control unobserved, spatially invariant factors, such as the El Niño or La Niña events. We also consider the linear region-specific time trends $h_{i}(t) = \lambda_{i}t$ to control gradual changes in individual regions’ growth rates driven by slowly changing factors, such as the gradual increased adaptation to the climate change.

Since the panel data cover decades, the coefficients in the panel model jointly capture the weather and climate effects. However, after including region-specific time trends, the impacts from gradual long-term climate change are expected to be absorbed. Therefore, the coefficients $\alpha$ and $\beta$ only capture the nonlinear effects of weather on output level, and the coefficients $\gamma_{0}$ and $\delta_{0}$ only capture the nonlinear effects of weather on output growth \citep{kalkuhl2020impact}. 

Due to varying definitions of subnational regions across countries, some countries have more granular subnational divisions, while others have coarser ones. For instance, Brazil and Italy have the same number of subnational regions, but Brazil is 28 times larger in area and has 3.6 times the population of Italy. Using subnational data directly, therefore, emphasizes the climate change responses of countries with more subnational divisions. To address this issue, we employ two strategies: First, we use the inverse of the number of subnational regions in a country as the weight in the regression (hereinafter referred to as region weighting). The interpretation of these results reflects the effects of climate on a country's average economic output, which allows us to compare our results with those from other studies based on country-level data. Second, we use the population of subnational regions as the weight (hereinafter referred to as population weighting). The interpretation of these results reflects the effect of climate on a person's average economic output (income). 

\subsection{Long-difference Model}
The long-difference model requires data over long periods. Previous studies generally employ a long-difference model with a single-period. However, this cross-sectional long-difference regression would be biased if within-region, time-varying factors are correlated with both climate and output \citep{burke2016adaptation}. To address this concern, we first take 5-year and 10-year averages of the data separately, that is, we set $m=5$ or $m=10$ in Equation (\ref{eq:emplongdiff}).
Then, we construct four-period and two-period panels of long difference models for regression based on Equation (\ref{eq:emplongdiff}):
\begin{equation}
\label{reg:longdiffemp}
\begin{aligned}
g_{i p}=\alpha \Delta \mathbf{T_{i p} T_{i p} }+\beta \Delta \mathbf{T_{i p}}+  \gamma \mathbf{T_{i p}^{2}}+\delta \mathbf{T_{i p}}+X_{i p}+\eta_{i}+\theta_{p}+h_{i}(p)+\epsilon_{i p}
\end{aligned}
\end{equation}
where $g_{i p}$ is the output per capita growth in region $i$ between period $p$ and period $p-1$. $\Delta \mathbf{T_{i p}}$ is the difference in the average temperature and precipitation of region $i$ between periods $p$ and $p-1$. Similar to the panel regression model, we also ignore the term of $\alpha_{0} \Delta \overline{T_{i p}} \cdot \overline{T_{i p-1}}$ to provide a parsimonious model and add it back in the robustness tests (Table \ref{tbl:altlag}). $X_{ip}$ are control variables, including regional urbanization rate, mean years of schooling, and population. Note that the long-difference model only eliminates the time-invariant factors relevant to the level of output (i.e. $c_{i}$), while time-invariant factors affecting the growth of output remain (i.e. $g_{i}$). To control the factors $g_{i}$, we include the region fixed effects $\eta_{i}$. $\theta_{p}$ is the period fixed effects. We also include the linear region-specific period trends $h_{i}(p) = \lambda_{i} p$ to control slowly changing factors over periods. 

Because of the longer time horizons in the long-difference model, the weather effects in this model can be expected to be rather small. Therefore, the $\alpha$ and $\beta$ capture the nonlinear effects of climate on output level, whereas the $\gamma$ and $\delta$ capture the nonlinear effects of climate on output growth. As with the panel regression, we also consider region weighting and population weighting in the long difference regression.

\section{Data and Descriptive Statistics}
\subsection{Data}
Temperature and precipitation data in this study are derived from the CRU database (https://crudata.uea.ac.uk/cru/data/hrg/). The CRU database offers a global, high-resolution (0.5\degree \texttimes 0.5\degree) monthly grids of land-based observations spanning from 1901 to the present. These data are developed based on station observations, interpolated onto grids using an angular-distance weighting method \citep{harris2020version}. This CRU database implements a degree of homogenization and shows no substantial discrepancies with other climate databases. It is widely used in the literature \citep{kalkuhl2020impact, song2023effects, malpede2024long}, allowing for the comparison of our results with previous findings.

To process the data, we first identify if a grid’s centroid lies within the boundaries of a region. The monthly grid data are then aggregated to the subnational level using area weights to obtain regions' monthly average temperature and monthly total precipitation. These monthly observations are subsequently aggregated\textemdash averaging for temperature and summing for precipitation\textemdash to derive annual mean temperatures and total precipitation at the subnational level.

Gross domestic product per capita (2011 PPP) data is obtained from \citet{kummu2018gridded}. The database was initially collected by \citet{gennaioli2013human} based on various government statistical agencies. It includes GDP data for 1569 subnational regions across 110 countries between 1990 and 2010, covering most countries in Central and South Africa that are less represented in other subnational GDP datasets. \citet{kummu2018gridded} extended the time series of this database from 2010 to 2015 and filled in missing countries based on national GDP data. Overall, the dataset developed by \citet{kummu2018gridded} provides global subnational GDP data from 1990 to 2015, without spatial gaps.

Population and urbanization data comes from the Global Historical Environment Database (HYDE 3.3) \citep{klein2017anthropogenic}. This dataset offers urban and rural population data at a 5 arc-minute $\times$ 5 arc-minute resolution from 10,000 BCE to 2023. Population data after 1950 is annual, sourced from the United Nations World Population Prospects and downscaled to a 5 arc-minute $\times$ 5 arc-minute grid based on the global 1 km × 1 km population grid data (LanSan) published by the Oak Ridge National Laboratory (USA). Additionally, this dataset classifies the global population into urban and rural populations based on urbanization rates, thereby providing the necessary population and urbanization data used in this study. To obtain the regions' population and urbanization, we first determine the grid's centroid, and then sum up the urban and rural gridded data into the subnational level if the grid's centroid falls within a region. The total population of a region is equal to the sum of the urban and rural populations, and the urbanization rate is calculated as the ratio of the urban population to the total population. The data of mean years of schooling comes from Global Data Lab (Access data: 22 November 2022)\footnote{https://globaldatalab.org/shdi}.

\subsection{Descriptive Statistics}
Table \ref{tbl: stats} summarizes the subnational data used in this study. Our sample includes 1,666 subnational regions from 196 countries, covering nearly all countries and populations worldwide (excluding control variables). Between 1990 to 2015, the global average GDP is \$11,315 per person per year. This is roughly equivalent to the average GDP per capita of Algeria and Thailand. Qatar has the highest average GDP per capita (\$104,617 per person per year), while Somalia has the lowest (\$607/p/yr). The global average subnational temperature is 18.6\celsius{} if weighted by regions, while the population-weighted temperature is 19.0\celsius. This slight increase in population-weighted temperature indicates that people tend to live in warmer regions. People also tend to live in drier regions, but the difference between average region-weighted precipitation (1.12m) and average population-weighted precipitation (1.11m) is relatively small. The average population per region is 5.7 million, with the highest being Uttar Pradesh in India, with a population of 220 million. The average urbanization and education level of regions are around 0.5 and 7 years, respectively.

\begin{figure}[!ht]
\includegraphics[width=\textwidth]{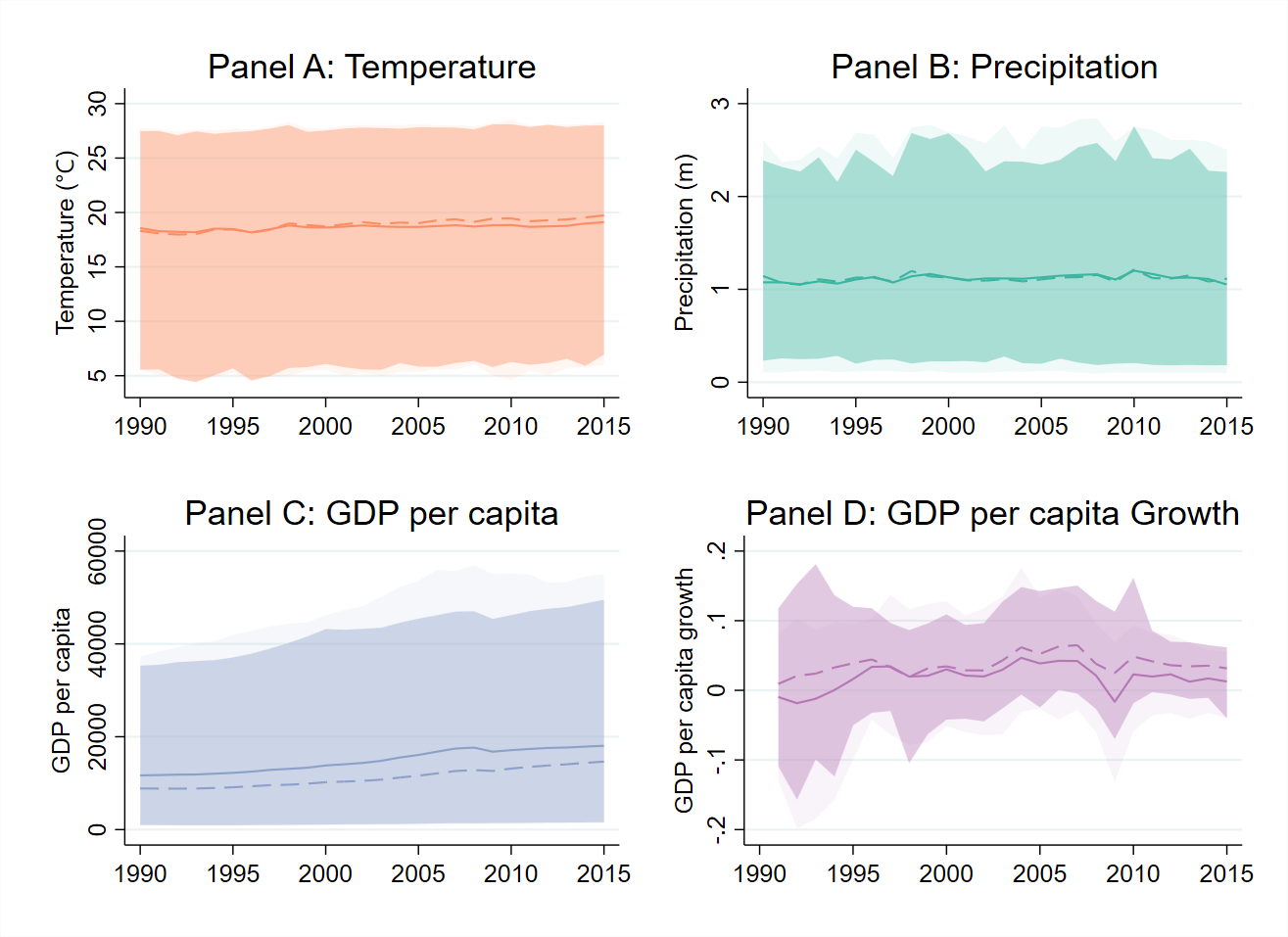}
\QTR{caption}{Changes in temperature, precipitation, and GDP per capita from 1990 to 2015.}
\label{fig: stats}
\begin{figurenotes}
This figure shows the global average temperature, precipitation, GDP per capita, and GDP per capita growth from 1990 to 2015. The solid lines represent the region-weighted average data. The dash lines represent the pop-weighted average data. Dark shadows represent the fifth to ninety-fifth percentile of region-weighted data. Light shadows represent the fifth to ninety-fifth percentile of population-weighted data.
\end{figurenotes}
\end{figure}

Figure \ref{fig: stats} shows the average temperature (Panel A), precipitation (Panel B), GDP per capita (Panel C), and GDP per capita growth rate (Panel D) over time. All these variables exhibit increasing trends starting from 1990. On average, the global temperature increased by 0.50\celsius, precipitation increased by 46 mm, GDP per capita increased by \$6,168, and the GDP per capita growth rate increased by 2.7\% when comparing the average values from 1990-1994 to those from 2011-2015. These increasing trends indicate an underlying non-stationary process, which may result in spurious results when panel models. In this case, we conduct a series of unit root tests, the results, however, suggest that all variables used for the annual panel model and the long-difference model are stationary (Table \ref{tbl:stationary}).

\begin{table}[!ht]
\caption{Summary statistic}
\label{tbl: stats}
\centering
\resizebox{\textwidth}{!}{
    \begin{tabular}{>{\raggedright}p{5.2cm}l c c c c c c c c c }
    \hline
        Variable & ~ & Mean & SD &Min & Max & Obs. & Regions & Countries & Year \\ \hline
        GDP per capita (population-weighted) & $y_{i t}$ & 11315 & 14141 & 177 & 459271.4 & \multirow{11}{*}{43316} & \multirow{11}{*}{1666} & \multirow{11}{*}{196} & \multirow{11}{*}{1990-2015} \\ 
        Annual mean temperature(\celsius) (region-weighted) & $T_{i t}$ & 18.64 & 8.14 & \multirow{2.5}{*}{-19.76} & \multirow{2.5}{*}{29.71} & ~ & ~ & ~ & ~ \\ 
        Annual mean temperature(\celsius) (population-weighted) & $T_{i t}$ & 18.96 & 7.39 & ~ & ~ & ~ & ~ & ~ & ~ \\ 
        Annual total precipitation(m) (region-weighted) & $P_{i t}$ & 1.12 & 0.77 & \multirow{2.5}{*}{0.00027} & \multirow{2.5}{*}{6.31} & ~ & ~ & ~ & ~ \\ 
        Annual total precipitation(m) (population-weighted) & $P_{i t}$ & 1.11 & 0.66 & ~ & ~ & ~ & ~ & ~ & ~ \\ 
        \hline
        Population (million) (region-weighted)& $pop_{i t}$ & 5.7 & 11.3 & 0.00048 & 222 & 43316 & 1666 & 196 &  \multirow{9}{*}{1990-2015} \\
        Urbanization (region-weighted)& $urb_{i t}$ & 0.53 & 0.24 &  \multirow{2.5}{*}{0} &  \multirow{2.5}{*}{1.0} &  \multirow{2.5}{*}{43215} &  \multirow{2.5}{*}{1663} &  \multirow{2.5}{*}{193} &  ~ \\ 
        Urbanization (population-weighted)& $urb_{i t}$ & 0.48 & 0.22 & ~ & ~ & ~ & ~ & ~ & ~ \\ 
        Education level(year) (region-weighted)& $edu_{i t}$ & 7.2 & 3.2 & \multirow{2}{*}{0.3} & \multirow{2}{*}{14.7} & \multirow{2}{*}{41470} & \multirow{2}{*}{1624} & \multirow{2}{*}{182} & ~ \\ 
        Education level(year) (population-weighted)& $edu_{i t}$ & 6.9 & 2.9 & ~ & ~ & ~ & ~ & ~ & ~\\ \hline
    \end{tabular}
}
\end{table}

\section{Empirical Results}
\subsection{Annual Panel Model}
Table \ref{tbl:panel} presents the annual panel regression results. Columns (1) to (4) are region-weighted results, while columns (5) and (6) use population weighting. Column (1) is based on the model developed by \citet{burke2015global}, which includes a quadratic function only capturing the aggregate effects of the weather conditions (sum effects of both the level and the growth). The results in column (1) show a significant effect of temperature on GDP per capita, while precipitation shows an insignificant effect. These findings align with \citet{burke2015global}, but the optimal temperature implied by column (1) is 15.8\celsius{}, 2.8\celsius{} higher than the value reported by \citet{burke2015global}. 

Column (2) presents the results based on Equation (\ref{reg: panelemp}), which captures the level and growth effects separately. We find that the effects of precipitation on both output level and growth are insignificant. The effect of temperature on output level is also insignificant. However, temperature has a significant effect on output growth. These results suggest that the significant temperature effect observed in column (1) is primarily driven by its effect on output growth rather than the output level. The optimal temperature implied by columns (2) is 14.6\celsius{}, but decreases to 11.6\celsius{} in columns (3) when we further include the control variables. The different optimal temperatures may be due to the control variables, but could also result from the reduced observations after including control variables. To investigate this further, we re-estimate column (2) using the same observations as in column (3). The result is shown in column (4). We find that the optimal temperature implied by column (4) is 11.8\celsius, closely matching the 11.6\celsius{} implied by column (3). This consistency indicates that the decrease in optimal temperature is primarily due to the reduced observations rather than the inclusion of control variables. To provide a representative global effect of temperature on output, we consider column (2) as our preferred specification, which includes more observations, and use it for the later analysis.

When using population weighting (columns 5 and 6), we find the results differ from those based on region weighting. Columns (5) and (6) show that the effects of temperature on both output level and growth are insignificant. However, there is a significant non-linear effect of precipitation on output level, but this effect is insignificant when precipitation equals 0 (the coefficient of $\Delta P$ is insignificant). To further explore the heterogeneity effects of weather conditions, we calculate the marginal effects of temperature and precipitation at different levels.

\begin{table}[!ht]
\QTR{caption}{Panel regression results}
\label{tbl:panel}
\centering
\resizebox{\textwidth}{!}{
    \begin{tabular}{>{\raggedright}p{2.5cm}l c c c c c c}
    \hline
        ~ & (1) & (2) & (3) & (4) & (5) & (6)  \\
        Dep. var. & \multicolumn{6}{c}{Annual GDP per capita growth} \\ \hline
        $\Delta T$ & ~ & -0.00588 & -0.00487 & -0.00514 & -0.000446 & -0.000613  \\ 
        ~ & ~ & (0.0046) & (0.0047) & (0.0047) & (0.0028) & (0.0030)  \\ 
        $\Delta T \cdot T$ & ~ & 0.000507 & 0.000560 & 0.000576 & 0.000190 & 0.000255  \\ 
        ~ & ~ & (0.0003) & (0.0004) & (0.0004) & (0.0002) & (0.0002)  \\ 
        $T$ & 0.0175*** & 0.0226** & 0.0168** & 0.0180** & 0.00358 & 0.00278  \\ 
        ~ & (0.0065) & (0.0094) & (0.0080) & (0.0082) & (0.0041) & (0.0039)  \\ 
        $T^2$ & -0.000554*** & -0.000774*** & -0.000725*** & -0.000760*** & -0.000163 & -0.000226  \\ 
        ~ & (0.0002) & (0.0003) & (0.0003) & (0.0003) & (0.0001) & (0.0001)  \\ 
        $\Delta P$ & ~ & -0.00288 & 0.00617 & 0.00639 & 0.0175 & 0.0196  \\ 
        ~ & ~ & (0.0094) & (0.0087) & (0.0086) & (0.0124) & (0.0130)  \\ 
        $\Delta P \cdot P$ & ~ & -0.000998 & -0.00474 & -0.00445 & -0.0127** & -0.0139**  \\ 
        ~ & ~ & (0.0039) & (0.0037) & (0.0036) & (0.0056) & (0.0063)  \\ 
        $P$ & 0.0134 & 0.0169 & 0.00592 & 0.00591 & 0.0130 & 0.00843  \\ 
        ~ & (0.0094) & (0.0145) & (0.0132) & (0.0130) & (0.0108) & (0.0122)  \\ 
        $P^2$ & -0.00448* & -0.00405 & -0.000674 & -0.00121 & -0.000692 & 0.000705  \\ 
        ~ & (0.0024) & (0.0035) & (0.0033) & (0.0033) & (0.0026) & (0.0032)  \\ \hline
        Obs. & 41650 & 41650 & 39915 & 39915 & 41650 & 39915  \\ 
        $R^2$ & 0.214 & 0.215 & 0.225 & 0.220 & 0.330 & 0.325  \\ 
        Region FE & Yes & Yes & Yes & Yes &Yes & Yes   \\
        Year FE & Yes & Yes & Yes & Yes  & Yes & Yes \\
        Region-specific time trends & Yes & Yes & Yes  & Yes & Yes & Yes  \\
        Control Var. & No & No & Yes & No & No & Yes \\
        Weight & Region & Region  & Region & Region &Pop. & Pop.    \\ \hline
    \end{tabular}
}
\begin{tablenotes}
Standard errors clustered at the country level are in parentheses. ***p \textless 0.01, **p \textless 0.05, *p \textless 0.10
\end{tablenotes}
\end{table}

\begin{figure}[!ht]
\includegraphics[width=\textwidth]{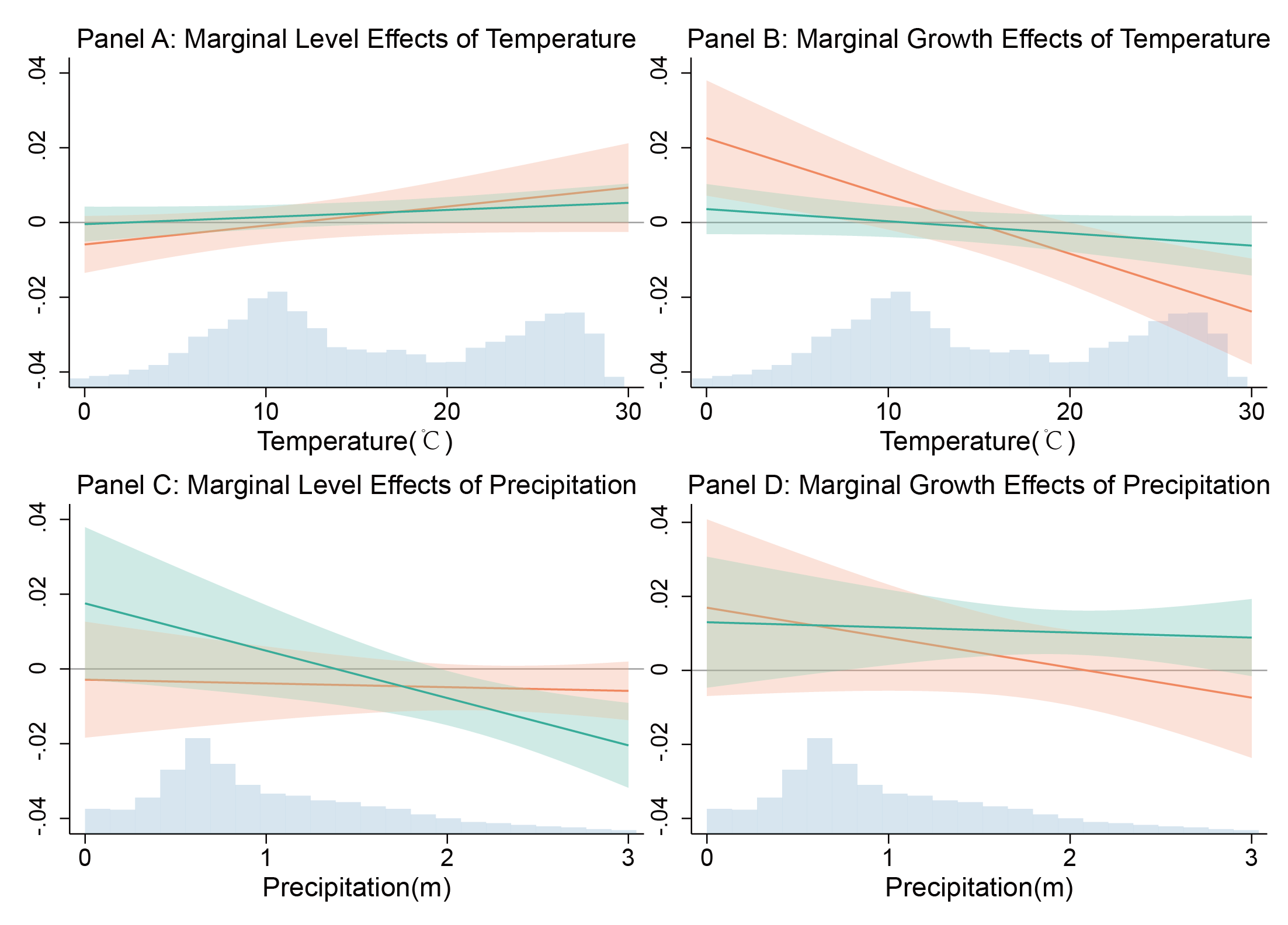}
\QTR{caption}{Marginal effects of temperature, precipitation on output\textemdash Panel estimates}
\label{fig:panelmarginal}
\begin{figurenotes}
This figure shows the marginal effects of temperature on GDP per capita growth (top), and the marginal effects of precipitation on GDP per capita growth (bottom) based on the panel model. The orange line represents the estimates based on region-weighted regression. The green line represents the estimates based on population-weighted regression. The shadow areas represent the 90\% confidence interval. Blue histograms are the distribution of temperature in Panel A and B, while they are the distribution of precipitation in Panel C and D.
\end{figurenotes}
\end{figure}

Figure \ref{fig:panelmarginal} shows the marginal effects of temperature and precipitation on output, derived from the region- and population-weighted results in columns (2) and (5) of Table \ref{tbl:panel}. When weighted by region, only temperature shows a significant marginal effect on output growth when temperatures are below 5\celsius{} or above 22\celsius{}. 1\celsius{} increase is expected to increase GDP per capita growth by 1.5\% in regions with an average temperature of 5\celsius{} and decrease GDP per capita growth by 1.6\% at 25\celsius{}. However, under population weighting, the marginal effects of temperature on output become insignificant. In contrast, precipitation consistently shows positive marginal effects on output growth. 100mm increase in precipitation is expected to increase GDP per capita growth by around 0.1\%. In addition, the precipitation also shows a significant negative marginal effect on output level in regions with annual total precipitation exceeding 2.1 meters (only observed in less than 10\% regions).

The difference between the region- and population-weighted results may be due to the fact that population-weighted results emphasize the responses of regions with higher populations. In populous regions, where domestic and industrial water demands are higher, economic development tends to be more sensitive to changes in precipitation. Consequently, an increase in precipitation generally has a positive impact on these areas. In contrast, regions with lower populations have lower water demand, thus, changes in precipitation have a limited impact on their output. However, for the effect of temperature, higher population density may enable regions to better adapt to temperature changes due to easier access to temperature control infrastructure (e.g. cities have more air conditioning than rural areas). Therefore, the population-weighted effect of temperature is less significant than the region-weighted effect.

\subsection{Long-difference Model}
Table \ref{tbl:longdiff} presents the panel long-difference regression results. Columns (1) to (4) show results weighted by region, while columns (5) and (6) are weighted by population. Columns (1) and (2) use five-year averages for the variables, while columns (3) to (6) are based on ten-year averages, as in \citet{kalkuhl2020impact}. We find that columns (1) and (2) show significant effects of both temperature and precipitation on output growth. The optimal temperature implied by column (2) is 14.8\celsius, consistent with the panel model findings. The optimal precipitation implied by column (2) is 1.7 m, which exceeds the annual total precipitation in more than 80\% of regions. However, when longer time intervals are considered, the significant effect of precipitation on output growth disappears (column 3), but the temperature effect remains significant. The optimal temperature implied by column (3) is 18.0\celsius, which is 3.4\celsius{} higher than that implied by the panel model. This effect remains significant, with a higher optimal temperature, when control variables are included in column (4). These results suggest that long-term climate adaptation reduces sensitivity to changes in precipitation and high temperatures.

Columns (5) and (6) present population-weighted results based on ten-year differences. In column (5), the effects of both temperature and precipitation are insignificant. However, after including control variables in column (6), the significance of temperature's effect on output growth increases, but is still only significant at 10\% (column 6). These results suggest that populous regions are less sensitive to temperature changes, consistent with the findings from the panel regression results.

\begin{table}[!ht]
\QTR{caption}{Long-difference regression results}
\label{tbl:longdiff}
\centering
\resizebox{\textwidth}{!}{
    \begin{tabular}{>{\raggedright}p{2.5cm}l c c c c c c}
    \hline
        ~ & (1) & (2) & (3) & (4) & (5) & (6)  \\
        Dep. var. & \multicolumn{6}{c}{Period average GDP per capita growth} \\ \hline
        $\Delta T $ & -0.172 & -0.172 & -0.351* & -0.478** & -0.00731 & -0.0355  \\ 
        ~ & (0.1153) & (0.1153) & (0.1839) & (0.2061) & (0.1089) & (0.1125)  \\
        $\Delta T \cdot T$ & 0.0117** & 0.0117** & 0.0146* & 0.0263*** & -0.00885 & -0.00327  \\ 
        ~ & (0.0051) & (0.0051) & (0.0084) & (0.0093) & (0.0062) & (0.0059)  \\ 
        $T$ & 0.387** & 0.387** & 0.534** & 0.731*** & 0.256 & 0.406*  \\ 
        ~ & (0.1693) & (0.1693) & (0.2355) & (0.2655) & (0.2085) & (0.2086)  \\ 
        $T^2$ & -0.0121*** & -0.0121*** & -0.0149*** & -0.0208*** & -0.00623 & -0.0106*  \\ 
        ~ & (0.0038) & (0.0038) & (0.0055) & (0.0057) & (0.0055) & (0.0059)  \\ 
        $\Delta P$ & -0.351 & -0.351 & 0.105 & 0.158 & 0.412 & 0.230  \\ 
        ~ & (0.2611) & (0.2611) & (0.4333) & (0.4213) & (0.3145) & (0.3250)  \\ 
        $\Delta P \cdot P$ & 0.229* & 0.229* & 0.0955 & 0.0862 & -0.252* & -0.219*  \\ 
        ~ & (0.1161) & (0.1161) & (0.2597) & (0.2502) & (0.1301) & (0.1236)  \\ 
        $P$ & 0.876 & 0.876 & 0.279 & -0.169 & 0.166 & 0.235  \\ 
        ~ & (0.5533) & (0.5533) & (0.6268) & (0.6010) & (0.6441) & (0.7486)  \\ 
        $P^2$ & -0.257** & -0.257** & -0.115 & 0.00309 & -0.0669 & -0.0540  \\ 
        ~ & (0.1219) & (0.1219) & (0.1122) & (0.1026) & (0.1068) & (0.1293)  \\ \hline
        Obs. & 6664 & 6664 & 3332 & 3242 & 3332 & 3242  \\ 
        $R^2$ & 0.720 & 0.720 & 0.680 & 0.697 & 0.861 & 0.864  \\ 
        Interval & 5 yr. & 5 yr. &10 yr.&10 yr. &10 yr. & 10 yr. \\
        Region FE & Yes & Yes & Yes & Yes &Yes & Yes   \\
        Year FE & Yes & Yes & Yes & Yes  & Yes & Yes \\
        Region-specific time trends & Yes & Yes & No  & No & No & No  \\
        Control Var. & No & No & No & Yes & No & Yes \\
        Weight & Region & Region  & Region & Region &Pop. & Pop.    \\ \hline
    \end{tabular}
}
\begin{tablenotes}
Standard errors clustered at the country level are in parentheses. ***p \textless 0.01, **p \textless 0.05, *p \textless 0.10
\end{tablenotes}
\end{table}

\begin{figure}[!ht]
\includegraphics[width=\textwidth]{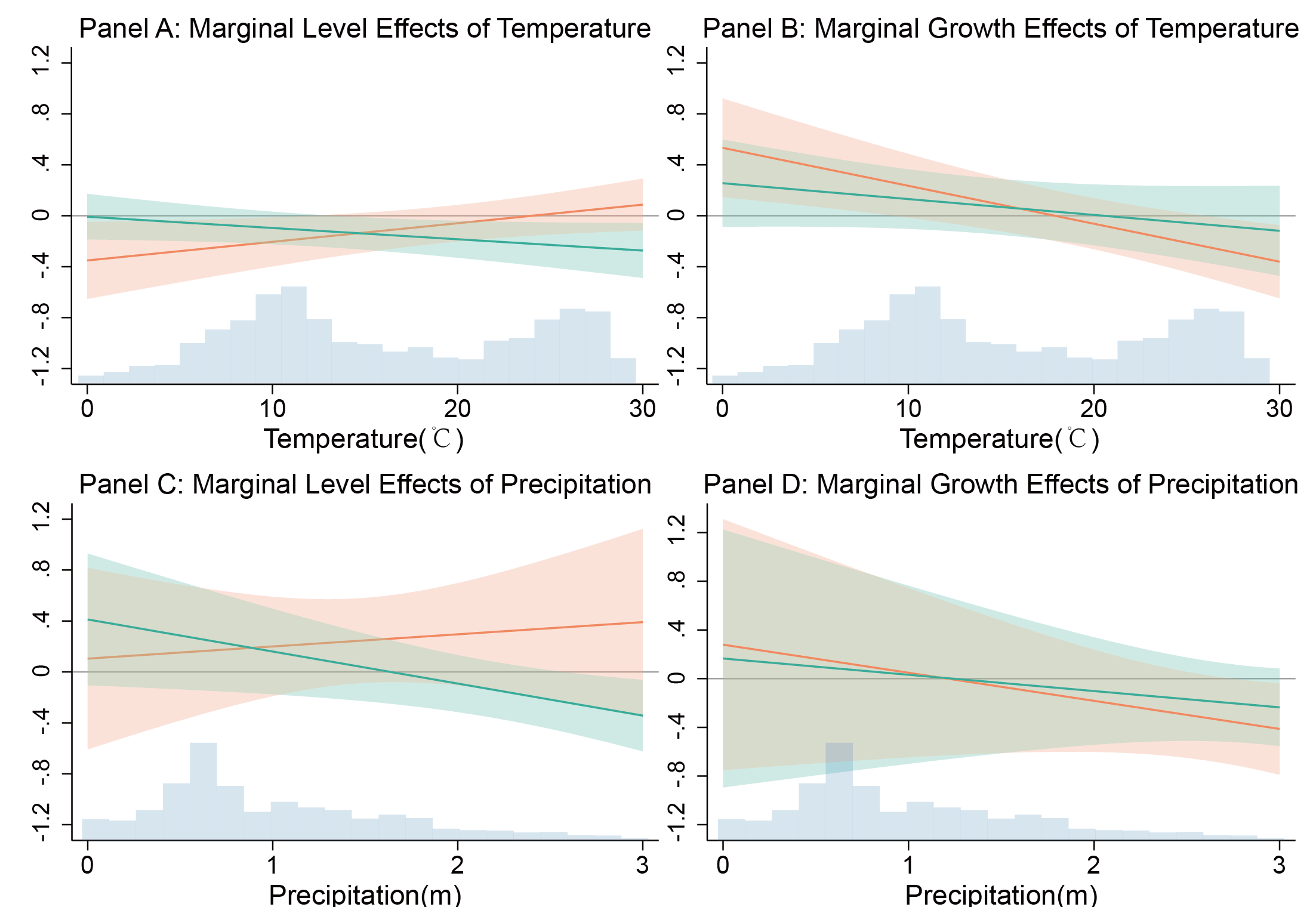}
\QTR{caption}{Marginal effects of temperature, precipitation on output \textemdash Long-difference estimates}
\label{fig:longdiffmarginal}
\begin{figurenotes}
This figure shows the marginal effects of temperature on GDP per capita growth (top), and the marginal effects of precipitation on GDP per capita growth (bottom) based on a long-difference model. The orange line represents the estimates based on region-weighted regression. The green line represents the estimates based on population-weighted regression. The shadow areas represent the 90\% confidence interval. Blue histograms are the distribution of temperature in Panel A and B, while they are the distribution of precipitation in Panel C and D.
\end{figurenotes}
\end{figure}

Figure \ref{fig:longdiffmarginal} shows the marginal effects of temperature and precipitation on output, based on region and population weighting from columns (3) and (5) in Table \ref{tbl:longdiff}. When weighted by region, the marginal effect of temperature on output level is insignificant (only significant at 10\% level for low-temperature regions). However, its marginal effect on output growth is significant, but this marginal effect is only significant (at 5\% or higher level ) at extreme cold and hot regions where the annual mean temperature is below 5\celsius{} or exceeds 29\celsius. A 1 \celsius{} increase in temperature is expected to increase GDP per capita growth between periods by 38.4\% in regions with an average temperature of 5 \celsius{} and decrease GDP per capita growth between periods by 33.1\% at 29\celsius{}. Since we consider a 10-year difference, these marginal effects indicate that a 1 \celsius{} increase in temperature is expected to increase annual GDP per capita growth by $\sqrt[10]{1+38.4\%}-1 = 3.3\%$ and decrease annual GDP per capita growth by $\sqrt[10]{1+33.1\%}-1 = 2.9\%$. In contrast, when weighted by population, the increase in temperature only decreases output level (Panel A) rather than output growth (Panel B). For instance, 1 \celsius{} increase at 20\celsius{} and 25\celsius{} decrease GDP per capita by 18.4\% and 22.9\%, respectively.

The marginal effects of precipitation on both output level and growth are insignificant when weighted by region or population. The sole exception is that 100mm increase in precipitation significantly decreases output level by 3.4\% at regions with annual total precipitation of 3m when weighted by population (Panel C), which is observed in less than 1\% regions in the world (mainly found in regions of less populated tropical rainforests in Colombia, Ecuador, Indonesia, Malaysia and Thailand). These findings indicate that precipitation has limited long-term effects on output, even in populous regions.

\subsection{Heterogeneity and Adaptation}
\label{sec:Heterogeneity And Adaptation}
In this section, we focus on the region-weighted effect of temperature on output growth, as it is the only significant effect in both the annual panel and the panel long-difference regression results. While population-weighted results indicate a significant effect of hot temperatures on output level in long-difference regression, these results emphasize populous regions, and are therefore less representative compared to the region-weighted results. The level effect is also expected to cause less damage compared to the growth effect in future output projections. To analyze the heterogeneity effects of temperature across different economic levels, we first calculate the 10-year average GDP per capita for each region to align with the time intervals of the long-difference model. We then classify the regions in each period as ``rich" or ``poor" based on the region-weighted median value. Regions with average GDP per capita above the median are classified as ``rich", while those below the median are classified as ``poor". Finally, we interact variables in equation (\ref{reg: panelemp}) and (\ref{reg:longdiffemp}) with a dummy variable representing the economic levels of regions\footnote{The panel model for heterogeneity analysis is: 

$g_{i t} = \alpha\mathbf{\Delta T_{i t} T_{i t}}+ \beta\mathbf{\Delta T_{i t}}+\gamma_{1} \mathbf{T^{2}_{i t}}+\delta_{1}\mathbf{T_{i t}}+poor \times (\gamma_{2} \mathbf{T^{2}_{i t}}+\delta_{2}\mathbf{T_{i t}})+X_{it}+\eta_{i}+\theta_{t}+h_{i}(t)+\epsilon_{i t}$. 
$poor$ is the dummy variable, which is equal to 0 for poor regions and equal to 1 for rich regions. The long-difference model is the same as the panel model, but the variables are averaged over 10 years}. 

\begin{figure}[!ht]
\includegraphics[width=\textwidth]{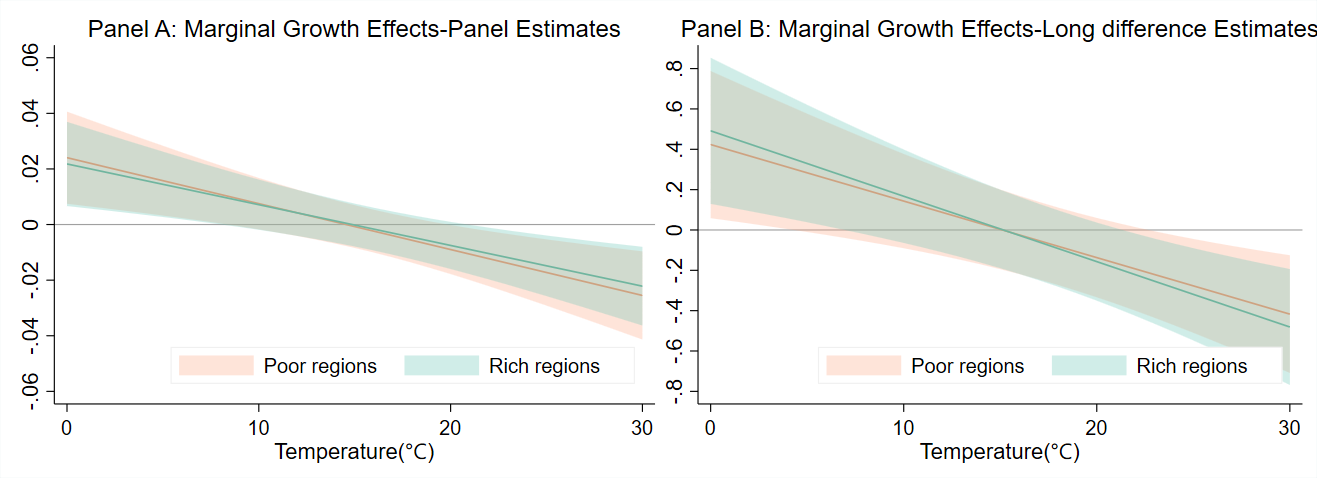}
\QTR{caption}{Marginal effects of temperature on output growth under different economic levels }
\label{fig:heterotem}
\begin{figurenotes}
This figure shows the marginal effects of temperature on GDP per capita growth based on the panel model (panel A) and the long-difference model (panel B). The orange line represents the poor regions, while the green line represents the rich regions. The shadow areas represent the 90\% confidence intervals.
\end{figurenotes}
\end{figure}

Panel A of Figure \ref{fig:heterotem} shows the marginal effects of temperature on output growth in rich and poor regions based on the annual panel model. We find that the marginal effect of temperature in poor regions is nearly identical to that in rich regions. 1 \celsius{} increase in temperature at 25 \celsius{} reduces GDP per capita growth by 1.7\% in poor regions and 1.5\% in rich regions. In addition, when using the panel long-difference model, the marginal effects of temperature are also consistent across rich and poor regions (Panel B). 1 \celsius{} increase in temperature at 25 \celsius{} is expected to reduce GDP per capita growth between periods by 27.7\% in rich regions and 31.9\% in poor regions (2.5\% and 2.8\% for annual GDP per capita growth). These results suggest that different economic regions have the same adaptation to weather and climate changes, and the vulnerability of poor regions to weather and climate change is primarily due to higher temperature rather than their economic status, consistent with the research of \citet{burke2015global} and \citet{,mendelsohn2006distributional}.

Apart from adaptation between regions, we also examine adaptation over time by comparing the marginal effects from the annual panel model and the panel long-difference model. As suggested by \citet{burke2016adaptation}, the annual panel model captures the short-run weather effects (denoted as $\hat{\tau}^{FE}$), while the long-difference model captures long-run climate effects (denoted as $\hat{\tau}^{LD}$). The value $1-\hat{\tau}^{LD}/\hat{\tau}^{FE}$, therefore, gives the percentage of the negative short-run effect that is offset in the longer run, measuring adaptation to climate change. To ensure comparability, we first convert the marginal growth effect of temperature between periods into the annual average marginal growth effect using $\hat{\tau}^{LD}=\sqrt[10]{1+\hat{\varphi}}-1$, where $\hat{\varphi}$ represents the marginal estimates based on the panel long-difference model. To quantify the uncertainty of the adaptation estimate, we further bootstrap our data 1000 times and calculate the $1-\hat{\tau}^{LD}/\hat{\tau}^{FE}$ for each iteration\footnote{Bootstrap estimates in Robustness checks section provide a detailed discussion about this approach}.

\begin{figure}[!ht]
\includegraphics[width=\textwidth]{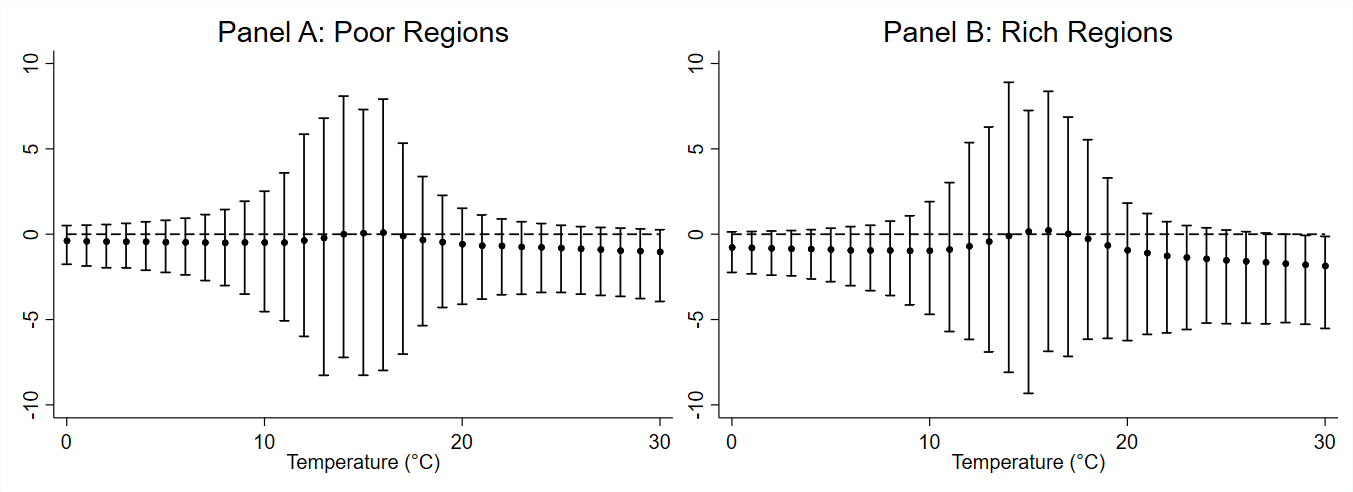}
\QTR{caption}{Marginal effects of temperature on output growth under different economic levels }
\label{fig:adaptation}
\begin{figurenotes}
This figure shows the percentage of the short-term effects of temperature on output growth that are mitigated in the longer run for poor (panel A) and rich (panel B) regions. The whiskers represent the fifth to ninety percentile.
\end{figurenotes}
\end{figure}

Figure \ref{fig:adaptation} presents the bootstrap results. Although the median estimates are negative at both low and high temperature levels, the confidence intervals are wide and include zero \textemdash except for the estimates where temperatures exceed 27\celsius{} in rich regions. These findings suggest that, for most regions, we cannot reject the hypothesis that sensitivity to short-run and long-run climate change is similar. However, in rich regions where average temperatures exceed 27\celsius{}, long-term damages are expected to be more severe, implying that long-term adaptation to temperature increases is still inadequate.

\section{Future Damage Projections}
Building on our annual panel and long-difference regression results, this section projects future output losses under a 2.0\celsius{} global warming scenario. We use the Shared Socioeconomic Pathway of Sustainability (SSP1) scenario as the trajectory for future population and GDP growth, and the Representative Concentration Pathway of 2.6 $W/m^2$ (RCP2.6) scenario as the future global warming scenario, as the SSP1-26 scenario is considered as the scenario aligns closely with the 2.0\celsius{} global warming target of the 2015 Paris Agreement. 
 
We also consider a probabilistic framework to account for uncertainties in the historical relationship between temperature and economic growth (uncertainty of damage function), as well as the spatial pattern of future mean annual temperature change associated with a given level of aggregate emissions (uncertainty of climate pattern), as suggested by \citet{burke2018large}. Specifically, we bootstrap our data 1000 times with annual panel and long-difference models to quantify the uncertainty of the damage function (We discuss this approach in detail in the Robustness Checks section). The uncertainty of climate patterns comes from 184 global climate simulations from 12 Earth system models provided by the sixth phase of the Coupled Model Intercomparison Project (CMIP6). This approach results in 184,000 possible output loss projections based on combinations of bootstrapped estimates and climate simulations.

For each bootstrap run $b$ and climate condition $c$, the percentage change in GDP per capita in each year $t$ for region $i$ is projected by the following equation:
\begin{equation}
\begin{aligned}
\Psi_{i t}^{b c} =ln(y_{i t}^{b c }(\phi_{i t}^{b c }))-ln(\tilde{y}_{i t}^{b c})=\sum_{s=1}^{t}\phi_{i s}^{b c }
\end{aligned}
\end{equation}
Where $\Psi_{i t}^{b c}$ is the percentage change in GDP per capita. $\tilde{y}_{i t}^{b c}$ is the counterfactual GDP per capita without temperature change. $\phi_{i t}^{b c}=g^{b}(T_{i t}^{c})- g^{b}(T_{i 0}^{c})$ is the additional estimated change in the GDP per capita growth $g$ due to the projected temperature increase above baseline climate $T_{i 0}$. The baseline climate $T_{i 0}$ is the average temperature from 2015 to 2019 for each region $i$. $g^{b}(T_{i t}^{c})=\hat{\alpha}^b (T^c)^2_{i t}+\hat{\beta}^b T_{i t}^c$ is the GDP per capita growth estimated based on our panel and long-difference damage functions for each bootstrap run $b$. For long-difference damage functions, $g^{b}(T_{i t}^{c})$ is further converted by $\sqrt[10]{1+g^{b}(T_{i t}^{c})}-1$ to get the annual GDP per capita growth. Section \ref{sec:Heterogeneity And Adaptation} suggests the damage functions between different economic regions are almost identical, therefore, we employ the same damage function based on equation (\ref{reg: panelemp}) and equation (\ref{reg:longdiffemp}) for each region. We apply two different weighting methods to derive global aggregated effects from regional estimates: population-weighted and GDP per capita weighted. These data come from the SSP1 scenarios provided by \citet{li2022spatiotemporal} and \citet{wang2022global}, which are the latest global gridded datasets. 

\begin{figure}[!ht]
\includegraphics[width=\textwidth]{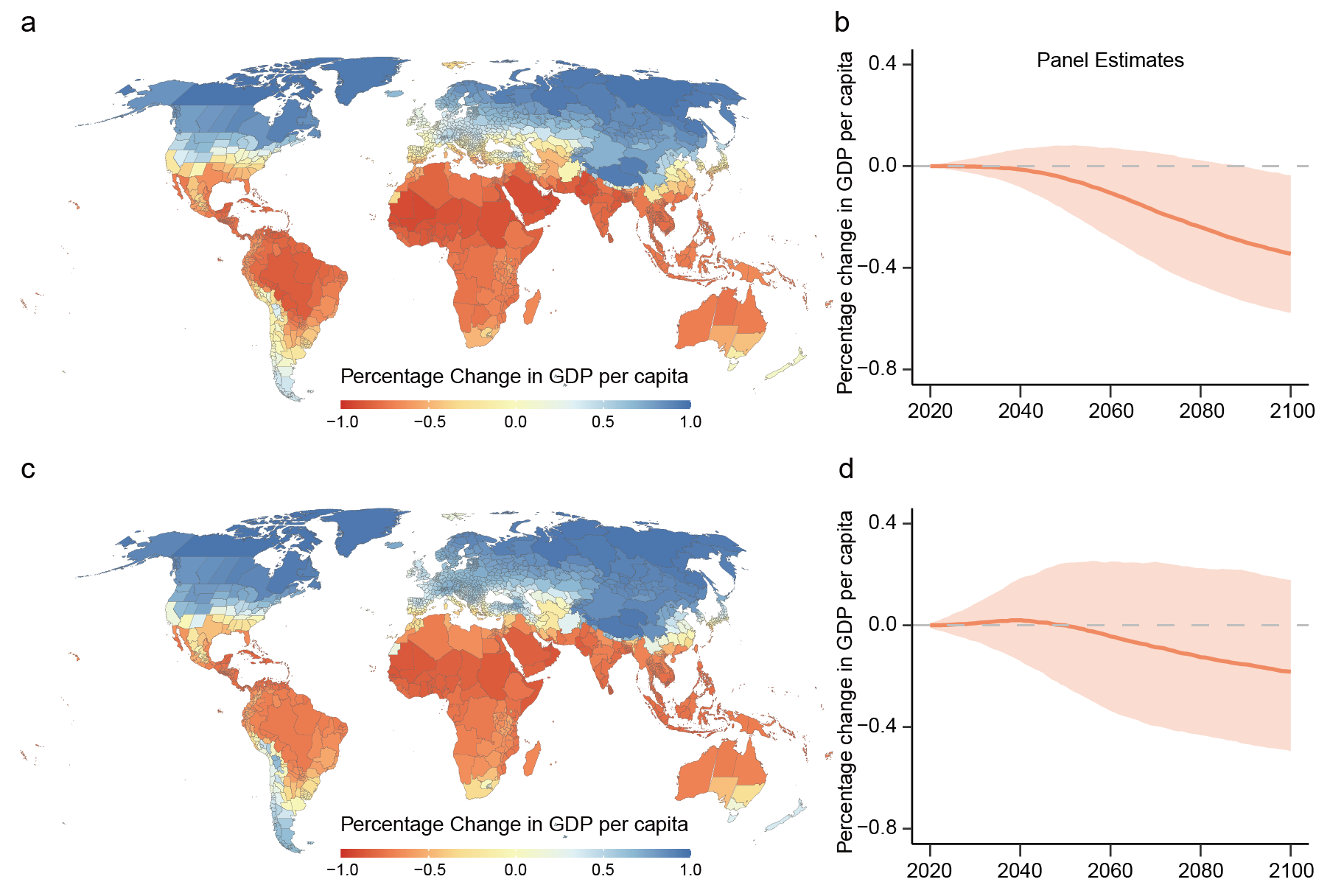}
\QTR{caption}{Regional and global aggregated GDP per capita changes}
\label{fig:loss}
\begin{figurenotes}
This figure shows the regional percentage changes in GDP per capita in 2100 (left) and global aggregated GDP per capita changes from 2020 to 2100 (right). A and b are the results based on the annual panel model, and c and d are the results based on the long-difference model. b,d are the GDPpc weighted results. 
\end{figurenotes}
\end{figure}

In practice, we randomly draw 1000 samples from bootstrapped estimates and climate simulations to calculate the uncertainty of the percentage change in GDP per capita\footnote{This sampling approach ensures a thorough sampling of the full parameter and scenario space and avoids computer memory issues}. Figure \ref{fig:loss} shows the projected regional GDP per capita changes in 2100 based on annual panel and long-difference models. Table \ref{tbl:loss} summarizes the GDP per capita changes at the subcontinent level. We find that temperature-induced output losses not only differ between countries, but also show pronounced variation within countries. For instance, while prior studies project temperature-related output declines for countries like the US and China\citep{burke2015climate,burke2018large}, our results suggest potential output increases in northern regions of these countries.

The annual panel model estimates indicate that the annual GDP per capita across regions will, on average, decline by 34.5–35.6\% in 2100 relative to a scenario with no additional warming after 2015–2019. This reduction is approximately three times larger than the findings of \cite{kalkuhl2020impact}, which suggest a level effect of temperature. In addition, the range between the 10th and 90th percentiles, weighted by population and GDP per capita, spans from -5.7\% to -60.0\% and -3.5\% to -57.7\%, respectively, with both ranges consistently below zero. This indicates high confidence in the negative effect of the increase in temperature if we consider short-term climate change (weather effect). However, these impacts vary widely by region: Sub-Saharan Africa, Southeast Asia, and Latin America are projected to face significant losses, while Europe and North America may see gains (Table \ref{tbl:loss}).

The long-difference estimates suggest milder impacts, with the regional average GDP per capita expected to decrease by 11.4–19.7\% on average in 2100, approximately one-third to one-half of the results based on annual panel estimates. In addition, the long-difference estimates show high uncertainty, allowing for potential positive effects of 22.0\% (population-weighted) and 17.8\% (GDPpc-weighted). These positive effects are primarily driven by the low temperatures in regions like Central and East Asia, Europe and North America, where their annual mean temperatures are lower than the optimal temperature. For example, the percentage change in Central and East Asia increases from 8.6\% to 48.8\% based on population-weighted annual panel and long-difference estimates.

\begin{table}[!ht]
\QTR{caption}{The percentage changes in GDP per capita at subcontinent level}
\label{tbl:loss}
    \centering
    \resizebox{\textwidth}{!}{
    \begin{tabular}{>{\raggedright}p{3.5cm} c c c c }
    \hline
        ~ & \multicolumn{2}{c}{Panel Estimates} & \multicolumn{2}{c}{Long-difference Estimates} \\ 
        ~ &  Population weighted & GDPpc weighted & Population weighted & GDPpc weighted \\ \hline
        Central and East Asia & 0.086 & -0.034 & 0.488 & 0.183  \\ 
        ~ & (0.329) & (0.25) & (0.473) & (0.397)  \\ 
        Southeast Asia & -0.857 & -0.875 & -0.833 & -0.819  \\ 
        ~ & (0.269) & (0.307) & (0.352) & (0.365)  \\ 
        South Asia & -0.823 & -0.663 & -0.491 & -0.491  \\ 
        ~ & (0.209) & (0.199) & (0.332) & (0.285)  \\ 
        Europe & 0.335 & 0.236 & 0.7 & 0.535  \\ 
        ~ & (0.315) & (0.295) & (0.42) & (0.407)  \\ 
        Middle East/North Africa & -0.722 & -0.585 & -0.499 & -0.361  \\ 
        ~ & (0.239) & (0.264) & (0.372) & (0.396)  \\ 
        Sub-Saharan Africa & -0.844 & -0.79 & -0.744 & -0.754  \\ 
        ~ & (0.238) & (0.24) & (0.335) & (0.325)  \\ 
        North America & 0.09 & 0.08 & 0.382 & 0.34  \\ 
        ~ & (0.303) & (0.271) & (0.419) & (0.376)  \\
        Latin America & -0.656 & -0.714 & -0.472 & -0.624  \\ 
        ~ & (0.243) & (0.251) & (0.317) & (0.33)  \\ 
          Global  &  -0.356 & -0.345  & -0.114 & -0.183  \\ 
           & (0.205) & (0.211) & (0.203) & (0.288)  \\ \hline
    \end{tabular}
    }
\begin{tablenotes}
Standard errors are in parentheses.
\end{tablenotes}
\end{table}

\section{Robustness Checks}
\textit{Data Source.}\textemdash We first compare our results with a global country-level output database \textemdash the World Bank database, which has also been widely used in previous studies. To ensure comparability, we aggregate our subnational data (Kummu database) to the country level. Table \ref{tbl:altoutput} shows the results. Columns (1) to (3) are the results based on the panel model, whereas columns (4) to (6) present the results based on the long-difference model. Column (1), which is based on the full set of observations, shows that only the growth effect of temperature is significant. This is consistent with the subnational level results in Table \ref{tbl:panel}. However, the significance and magnitude of the effect are lower than that in column (3) from Table \ref{tbl:panel}. The optimal temperature implied by column (1) is 16.2 \celsius{}, which is 1.6 \celsius{} higher than the value derived from subnatioanl-level data. These results confirm the findings of \citet{damania2020does}, which suggests that the data aggregating to large spatial scales masks the some heterogeneity impact of climate, thus underestimating the impact of climate. 

Columns (2) and (3) are the results based on the Kummu and World Bank databases for their shared countries, respectively. The results suggest that the coefficients in both columns are nearly identical, although the significance of temperature effects is reduced compared to column (1), as one might expected given the smaller sample. Columns (4) to (6), which are based on the long-difference model, also show a significant effect of temperature on output growth with a reduced magnitude compared to the subnational level results in Table \ref{tbl:longdiff}. The magnitude and significance of the temperature coefficients are also consistent across columns. These consistences suggest the reliability of the Kummu database used in this study.

\begin{table}[!ht]
\QTR{caption}{Regression results with alternative country-level output database}
\label{tbl:altoutput}
    \centering
    \resizebox{\textwidth}{!}{
    \begin{tabular}{>{\raggedright}p{2.5cm} c c c c c c}
    \hline
        ~ & (1) & (2) & (3) & (4) & (5) & (6) \\ 
        Data source & Kummu & Kummu& World Bank& Kummu & Kummu & World Bank \\
        Model &  \multicolumn{3}{c}{Panel model} &  \multicolumn{3}{c}{Long-differece model} \\ \hline
         $\Delta T$ & -0.00139 & 0.000654 & 0.00178 & -0.237 & -0.241 & -0.291  \\
        ~ & (0.0049) & (0.0048) & (0.0049) & (0.2112) & (0.2102) & (0.1927)  \\ 
        $\Delta T \cdot T$ & 0.000165 & 0.0000256 & -0.000109 & 0.00799 & 0.00699 & 0.0115  \\ 
        ~ & (0.0003) & (0.0003) & (0.0003) & (0.0097) & (0.0097) & (0.0089)  \\ 
        $T$ & 0.0179* & 0.0149 & 0.0162 & 0.488* & 0.531** & 0.501**  \\ 
        ~ & (0.0099) & (0.0098) & (0.0099) & (0.2523) & (0.2508) & (0.2513)  \\ 
        $T^2$ & -0.000554** & -0.000466* & -0.000518* & -0.0133** & -0.0133** & -0.0119**  \\ 
        ~ & (0.0003) & (0.0003) & (0.0003) & (0.0059) & (0.0058) & (0.0057)  \\ 
        $\Delta P$ & -0.00446 & -0.000404 & -0.000687 & 0.0895 & 0.195 & 0.114  \\ 
        ~ & (0.0120) & (0.0120) & (0.0120) & (0.5601) & (0.5659) & (0.5032)  \\ 
        $\Delta P \cdot P$ & -0.000187 & -0.00152 & -0.00229 & 0.107 & 0.0595 & 0.0830  \\ 
        ~ & (0.0050) & (0.0050) & (0.0052) & (0.3294) & (0.3318) & (0.2999)  \\ 
        $P$ & 0.0207 & 0.00980 & 0.0103 & 0.874 & 0.521 & 0.402  \\ 
        ~ & (0.0197) & (0.0188) & (0.0171) & (0.8156) & (0.8309) & (0.7886)  \\ 
        $P^2$ & -0.00575 & -0.00367 & -0.00260 & -0.219 & -0.148 & -0.110  \\ 
        ~ & (0.0048) & (0.0046) & (0.0043) & (0.1499) & (0.1497) & (0.1356)  \\ \hline
        Obs. & 4925 & 4581 & 4581 & 394 & 362 & 362  \\ 
        $R^2$ & 0.258 & 0.256 & 0.256 & 0.705 & 0.681 & 0.690  \\ 
        Interval & 1 yr. & 1 yr. &1 yr.&10 yr. &10 yr. & 10 yr. \\
        Region FE & Yes & Yes & Yes & Yes &Yes & Yes   \\
        Year FE & Yes & Yes & Yes & Yes  & Yes & Yes \\
        Region-specific time trends & Yes & Yes & Yes  & No & No & No  \\
        Control Var. & No & No & No & No & No & No \\
        Weight & Region & Region  & Region & Region &Region & Region    \\ \hline
    \end{tabular}
    }
    \begin{tablenotes}
Standard errors clustered at the country level are in parentheses. ***p \textless 0.01, **p \textless 0.05, *p \textless 0.10
\end{tablenotes}
\end{table}

Second, we compare our results with another subnational database collected by \citet{kalkuhl2020impact} (henceforth the Kalkuhl database). The Kalkuhl database includes output data for 1518 subnational regions across 77 countries from 1900 to 2014. To ensure consistency, we limit the time series for both the Kummu and Kalkuhl databases to the period from 1990 to 2014 and apply region weighting for each regression. Table \ref{tbl:altsuboutput} shows the regression results. Column (1) is the results based on the Kummu database with all observations. The results show a significant effect of temperature on output growth, which is consistent with our main findings in Table \ref{tbl:panel}, although the magnitude of the coefficients is slightly larger. However, when we restrict the countries in the Kummu database to match those in the Kalkuhl database, the effect of temperature on output growth becomes insignificant (column 2). This is consistent with the result from the Kalkuhl database (column 3), confirming that incomplete data tends to underestimate the climate effects.

The long-difference regression results further confirm the bias estimates induced by the incomplete data. When we use the full set of observations from the Kummu database, we find significant effects of temperature on both output level and growth (column 4). However, when we limit the countries in the Kummu database to align with the Kalkuhl database, the significance of temperature on both output level and growth disappeared (column 5). Overall, the Kalkuhl database omits many hot and poor countries in Africa, Southeast Asia and Central America. The average temperature and GDP per capita from 1991 to 2014 for the 77 countries included in the Kalkuhl database are 10.4 \celsius{} and \$12334 (median value: 10.4 \celsius{} and \$5535), while they are 13.6 \celsius{} and \$11194 for 196 countries (median value: 17.9 \celsius{} and \$4974). Using incomplete data, therefore, is expected to underestimate the global climate effects.

\begin{table}[!ht]
\QTR{caption}{Regression results with alternative subnational level output database}
\label{tbl:altsuboutput}
    \centering
    \resizebox{\textwidth}{!}{
    \begin{tabular}{>{\raggedright}p{2.5cm} c c c c c c}
    \hline
        ~ & (1) & (2) & (3) & (4) & (5) & (6) \\ 
        Data source & Kummu & Kummu& Kalkuhl& Kummu & Kummu & Kalkuhl \\
        Model &  \multicolumn{3}{c}{Panel model} &  \multicolumn{3}{c}{Long-differece model} \\ \hline
         $\Delta T$ & -0.00733 & 0.00615 & -0.00210 & -0.554*** & 0.00112 & 0.00803  \\ 
        ~ & (0.0058) & (0.0091) & (0.0129) & (0.1780) & (0.0521) & (0.1218)  \\
        $L.\Delta T$ & 0.000706 & 0.00123 & 0.00158 & ~ & ~ &   \\ 
        ~ & (0.0034) & (0.0033) & (0.0065) & ~ & ~ &   \\
        $\Delta T \cdot T$ & 0.000636 & 0.000586 & -0.000304 & 0.0252*** & 0.000825 & -0.00948  \\ 
        ~ & (0.0004) & (0.0008) & (0.0011) & (0.0083) & (0.0055) & (0.0070)  \\
        $L.\Delta T \cdot T$ & -0.0000845 & 0.000299 & -0.000940 & ~ & ~ &   \\
        ~ & (0.0003) & (0.0004) & (0.0006) & ~ & ~ &   \\
        $T$ & 0.0251** & -0.00953 & 0.0122 & 0.597** & -0.00297 & 0.00346  \\
        ~ & (0.0107) & (0.0139) & (0.0198) & (0.2653) & (0.0024) & (0.0038)  \\
        $T^2$ & -0.000864*** & -0.000260 & -0.000270 & -0.0147** & 0.0000449 & -0.000131  \\
        ~ & (0.0003) & (0.0006) & (0.0008) & (0.0061) & (0.0001) & (0.0001)  \\
        $\Delta P$ & 0.0119 & 0.0263* & 0.0599 & 0.277 & 0.189 & -0.718**  \\ 
        ~ & (0.0129) & (0.0142) & (0.0505) & (0.3624) & (0.1667) & (0.3273)  \\ 
        $L.\Delta P$ & 0.00656 & 0.00920 & 0.0599** & ~ & ~ &   \\
        ~ & (0.0087) & (0.0068) & (0.0289) & ~ & ~ &   \\ 
        $\Delta P \cdot P$ & -0.00631 & -0.0147** & -0.0234 & 0.0163 & -0.111 & 0.116  \\ 
        ~ & (0.0058) & (0.0059) & (0.0227) & (0.1759) & (0.0689) & (0.1404)  \\ 
        $L.\Delta P \cdot P$ & -0.00232 & -0.00285 & -0.0223* & ~ & ~ &   \\ 
        ~ & (0.0038) & (0.0029) & (0.0120) & ~ & ~ &   \\  
        $P$ & 0.00295 & -0.0209 & -0.0688 & 0.429 & -0.0732* & -0.0419  \\ 
        ~ & (0.0182) & (0.0192) & (0.0656) & (0.5977) & (0.0379) & (0.0475)  \\ 
        $P^2$ & -0.00121 & 0.00473 & 0.0166 & -0.154 & 0.0191** & 0.0138  \\ 
        ~ & (0.0043) & (0.0039) & (0.0146) & (0.1139) & (0.0095) & (0.0127)  \\ \hline
        Obs. & 38318 & 22586 & 24427 & 3332 & 862 & 1306  \\ 
        $R^2$ & 0.241 & 0.261 & 0.432 & 0.694 & 0.532 & 0.668  \\ 
        Interval & 1 yr. & 1 yr. &1 yr.&10 yr. &10 yr. & 10 yr. \\
        No.countries & 196 & 77 & 77 & 196 & 63 & 63 \\
        Region FE & Yes & Yes & Yes & Yes &Yes & Yes   \\
        Year FE & Yes & Yes & Yes & Yes  & Yes & Yes \\
        Region-specific time trends & Yes & Yes & Yes  & No & No & No  \\
        Control Var. & No & No & No & No & No & No \\
        Weight & Region & Region  & Region & Region &Region & Region    \\ \hline
    \end{tabular}
    }
    \begin{tablenotes}
Standard errors clustered at the country level are in parentheses. ***p \textless 0.01, **p \textless 0.05, *p \textless 0.10
\end{tablenotes}
\end{table}

Third, we replace our climate database on the CRU database with the ERA5 database. The ERA5 database is a reanalysis climate database that provides global gridded, homogenized climate data. Compared to the CRU database, which relies on observations from weather stations, reanalysis data can compensate for limitations associated with regions where station observations spare (e.g. Africa). Table \ref{tbl:altclimate} shows the regression results. Column (2) is the results based on ERA5. It shows a significant effect of temperature on output growth, consistent with the results based on the CRU database (column 1), although the coefficient values for temperature in column (2) are slightly smaller than those in column (1). 

Columns (3) and (4), based on the long-difference model, also show a significant effect of temperature on output growth when using CRU and ERA5 data. However, column (4) shows significant effects of precipitation on both output level and growth, which is absent in column (3). The optimal precipitation for output level is 2.0m, which is higher than the annual total precipitation in over 90\% regions. This suggests that an increase in precipitation is expected to increase the output level in most regions. In contrast, the increase in precipitation continues to have a negative effect on output growth, as the relationship between precipitation and output growth is convex, with the precipitation corresponding to the vertex of the curve at 3.0m. Overall, the effects of precipitation are not robust, as they are only significant when using ERA5 data. In contrast, the effect of temperature on output growth is consistently significant when using both CRU and ERA5 data. Therefore, this study focuses on the temperature effects only. 

\begin{table}[!ht]
\QTR{caption}{Regression results with alternative climate database}
\label{tbl:altclimate}
    \centering
    \resizebox{\textwidth}{!}{
    \begin{tabular}{>{\raggedright}p{2.5cm} c c c c}
    \hline
        ~ & (1) & (2) & (3) & (4)  \\ 
        Data source & CRU & ERA & CRU & ERA  \\
        Model &  \multicolumn{2}{c}{Panel model} &  \multicolumn{2}{c}{Long-differece model} \\ \hline
         $\Delta T$ & -0.00588 & -0.00314 & -0.351* & -0.192*  \\
        ~ & (0.0046) & (0.0041) & (0.1839) & (0.1148)  \\ 
        $\Delta T \cdot T$ & 0.000507 & 0.000454 & 0.0146* & 0.00931*  \\ 
        ~ & (0.0003) & (0.0003) & (0.0084) & (0.0050)  \\ 
        $T$ & 0.0226** & 0.0156** & 0.534** & 0.508**  \\ 
        ~ & (0.0094) & (0.0074) & (0.2355) & (0.2021)  \\ 
        $T^2$ & -0.000774*** & -0.000685*** & -0.0149*** & -0.0130***  \\ 
        ~ & (0.0003) & (0.0002) & (0.0055) & (0.0049)  \\ 
        $\Delta P$ & -0.00288 & -0.00641 & 0.105 & 0.690***  \\ 
        ~ & (0.0094) & (0.0048) & (0.4333) & (0.1860)  \\
        $\Delta P \cdot P$ & -0.000998 & 0.000606 & 0.0955 & -0.169***  \\ 
        ~ & (0.0039) & (0.0008) & (0.2597) & (0.0557)  \\ 
        $P$ & 0.0169 & 0.00134 & 0.279 & -1.037**  \\ 
        ~ & (0.0145) & (0.0072) & (0.6268) & (0.4047)  \\ 
        $P^2$ & -0.00405 & -0.000265 & -0.115 & 0.170***  \\ 
        ~ & (0.0035) & (0.0007) & (0.1122) & (0.0643)  \\ \hline
        Obs. & 39984 & 39984 & 3332 & 3332   \\ 
        $R^2$ & 0.215 & 0.215 & 0.680 & 0.693   \\ 
        Interval & 1 yr. & 1 yr. &10 yr.&10 yr.  \\
        Region FE & Yes & Yes & Yes & Yes    \\
        Year FE & Yes & Yes & Yes & Yes   \\
        Region-specific time trends & Yes & Yes & No  & No   \\
        Control Var. & No & No & No & No  \\
        Weight & Region & Region  & Region & Region    \\ \hline
    \end{tabular}
    }
    \begin{tablenotes}
Standard errors clustered at the country level are in parentheses. ***p \textless 0.01, **p \textless 0.05, *p \textless 0.10
\end{tablenotes}
\end{table}

\textit{Nonlinear Effects.}\textemdash Following previous studies, we employ a quadratic function to capture the relationship between temperature and output. However, some studies also suggest linear or piecewise linear relationships. To explore this further, we use a non-parametric method to analyze the relationship between temperature and output. Specifically, we estimate the following regression:
\begin{equation}
\label{reg:baselineemp}
\begin{aligned}
g_{i t} =\sum_{m} \alpha^{m} \tilde{T_{i t}^{m}} + \sum_{n} \beta^{n} \tilde{P_{i t}^{n}} +\eta_{i}+\theta_{t}+h_{i}(t)+\epsilon_{i t}
\end{aligned}
\end{equation}
where $g_{i t}$ is the annual GDP per capita growth in region $i$ and year $t$. $\tilde{T_{i t}^{m}}$ is the annual mean temperature in region $i$ and year $t$ that belongs to $m$th temperature bin. Each interior temperature bin is 3\celsius{} wide. We define $\tilde{T}_{i t}^{m=0}$ when $T_{i t}<0$. The top ($m=11$) bin counts temperature $\geq 27$. $P_{i t}^{n}$ is defined similarly for annual total precipitation across 12 bins. Each bin spans 0.2m with the top bin corresponding to precipitation $\geq 2.2m$. $\eta_{i}$ is the region fixed effects.  $\theta_{t}$ is the year fixed effects. $h_{i}(t)$ is the linear region-specific time trend fixed effects. $\alpha^{m}$ and $\beta^{n}$ are the parameters of interest. They capture the aggregate marginal effects of temperature and precipitation on output in each bin $m$ and $n$. 

\begin{figure}[!ht]
\includegraphics[width=\textwidth]{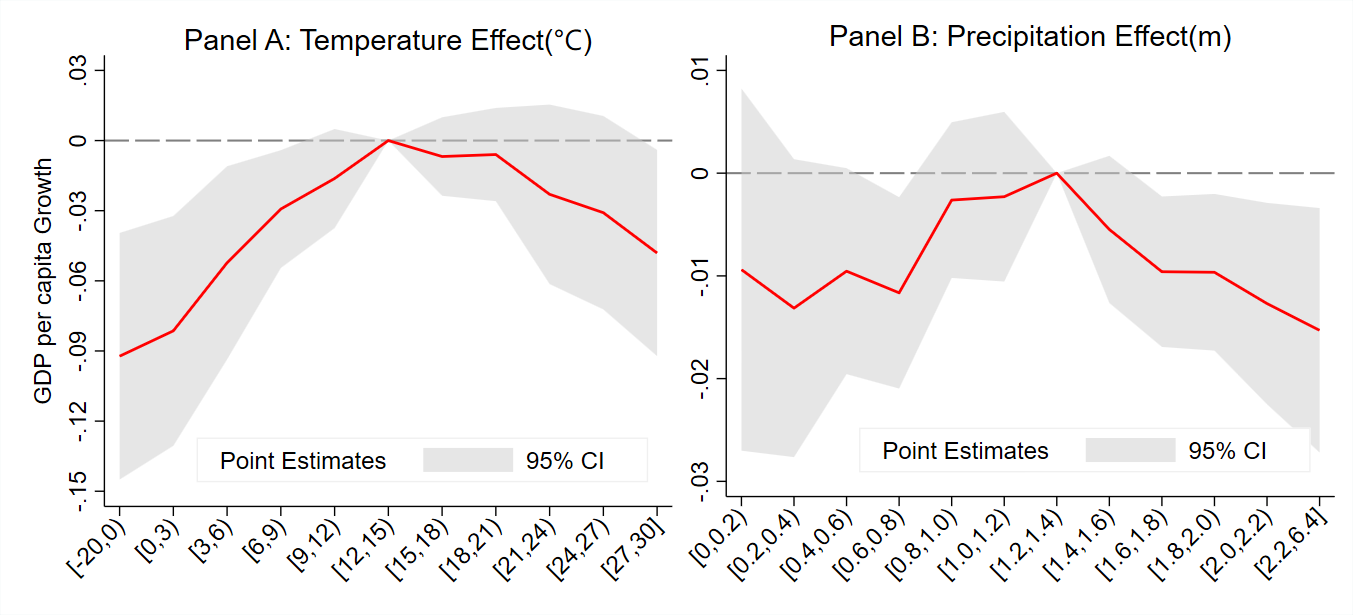}
\QTR{caption}{Point estimate results for temperature and precipitation}
\label{fig:baseline}
\begin{figurenotes}
This figure shows the effects of temperature (Panel A) and precipitation (Panel B) on GDP per capita growth based on baseline regression. Both of them are based on region-weighted regression. The shadow areas represent the 95\% confidence interval. 
\end{figurenotes}
\end{figure}

Figure \ref{fig:baseline} shows the baseline regression results based on equation (\ref{reg:baselineemp}). Panel A shows a quadratic relationship between temperature and GDP per capita growth. The optimal temperature for GDP per capita growth is around 12-15\celsius, slightly lower than that implied by column (1) in Table \ref{tbl:panel} (15.8\celsius). Lower temperature significantly reduces GDP per capita growth, while higher temperature only shows a significant negative effect in extremely hot regions (temperature $\geq 27\celsius$).

Precipitation exhibits a similar quadratic relationship with GDP per capita. The optimal precipitation is around 1.2-1.4m, slightly lower than the value implied by column (1) in Table \ref{tbl:panel} (1.5m). However, the relationship between GDP per capita and low precipitation is more likely to be linear (especially when precipitation is below 0.8m), with large confidence intervals. This may explain the weak significance (only significant at the 10\% level) observed when using the quadratic function to capture the relationship between precipitation and GDP per capita in column (1) of Table \ref{tbl:panel}. Overall, the point estimates support the validity of using the quadratic function to capture the effects of temperature and precipitation on output. However, for precipitation, a piecewise function might provide a better fit for analyzing its effects on output, which is beyond the scope of this study but deserves further research. 

\textit{Alternative Specifications.}\textemdash Table \ref{tbl:altspec} presents the effects of temperature and precipitation on output based on different fixed effects. Columns (1) to (3) are the annual panel regression results. Column (1) is our main specification that considers subnational region, year fixed effects and linear region-specific time trends. Column (2) replaces the linear region-specific time trends with the quadratic region-specific time trends to control for nonlinear gradual changes. Column (3) further includes the continent-by-year fixed effects to provide more flexible control over time trends across the continent. The term ``continent" here refers to 22 subcontinent regions. The results in columns (1) to (3) show that the effect of temperature on output growth is consistently significant across all specifications. The coefficient values and standard errors in columns (1) and (2) are almost identical, and slightly increased after considering continent-by-year fixed effects in columns (3).

Columns (4) and (5) are the long-difference regression results. Column (4) is our main specification that considers subnational region and year fixed-effects. Column (5) further includes the continent-by-year fixed effects. The results show that the effect of temperature on output growth remains consistently significant, with similar coefficient values across both specifications. In addition, we find that the effect of temperature on output level also becomes significant in column (5). However, in contrast to the findings of \citet{kalkuhl2020impact}, which reported a negative effect of temperature increases on output level, we observe a positive effect. The temperature corresponding to the vertex is 7.9\celsius{}. 1\celsius{} in temperature increase is expected to decrease GDP per capita by 22.1\% at 5\celsius{} regions and increase GDP per capita by 18.5\% at 25\celsius{}, but these effects are significant at 10\%. The marginal effects are statistically significant at the 5\% level only in extreme temperature regions, where the annual mean temperature exceeds 27\celsius{} or falls below 2\celsius{}.

\begin{table}[!ht]
\QTR{caption}{Regression results with different fixed effects}
\label{tbl:altspec}
    \centering
    \resizebox{\textwidth}{!}{
    \begin{tabular}{>{\raggedright}p{2.5cm} c c c c  c}
    \hline
        ~ & (1) & (2) & (3) & (4) & (5)  \\ 
        Model &  \multicolumn{3}{c}{Panel model} &  \multicolumn{2}{c}{Long-differece model} \\ \hline
         $\Delta T$  & -0.00588 & -0.00588 & -0.00552 & -0.351* & -0.322**  \\ 
        ~ & (0.0046) & (0.0047) & (0.0054) & (0.1839) & (0.1543)  \\
        $\Delta T \cdot T$ & 0.000507 & 0.000507 & 0.000302 & 0.0146* & 0.0203***  \\ 
        ~ & (0.0003) & (0.0003) & (0.0003) & (0.0084) & (0.0074)  \\ 
        $T$ & 0.0226** & 0.0226** & 0.0203* & 0.534** & 0.471*  \\
        ~ & (0.0094) & (0.0096) & (0.0120) & (0.2355) & (0.2537)  \\ 
        $T^2$ & -0.000774*** & -0.000774*** & -0.000828** & -0.0149*** & -0.0161***  \\ 
        ~ & (0.0003) & (0.0003) & (0.0003) & (0.0055) & (0.0061)  \\ 
        $\Delta P$ & -0.00288 & -0.00288 & -0.000837 & 0.105 & 0.245  \\ 
        ~ & (0.0094) & (0.0096) & (0.0100) & (0.4333) & (0.4496)  \\
        $\Delta P \cdot P$ & -0.000998 & -0.000998 & -0.00236 & 0.0955 & 0.103  \\ 
        ~ & (0.0039) & (0.0040) & (0.0042) & (0.2597) & (0.2758)  \\
        $P$ & 0.0169 & 0.0169 & 0.0214 & 0.279 & 0.492  \\ 
        ~ & (0.0145) & (0.0148) & (0.0141) & (0.6268) & (0.6005)  \\ 
        $P^2$ & -0.00405 & -0.00405 & -0.00406 & -0.115 & -0.193*  \\ 
        ~ & (0.0035) & (0.0036) & (0.0035) & (0.1122) & (0.1111)  \\ \hline
        Obs. & 41650 & 41650 & 41650 & 3332 & 3332   \\ 
        $R^2$ & 0.215 & 0.215 & 0.298 & 0.680 & 0.740 \\ 
        Interval & 1 yr. & 1 yr. &1 yr.&10 yr. &10 yr. \\
        Region FE & Yes & Yes & Yes & Yes &Yes    \\
        Year FE & Yes & Yes & Yes & Yes  & Yes  \\
        Region-specific time trends & Linear & Quadratic & Quadratic  & No & No  \\
        Continent by year FE & No & No & Yes  & No & Yes  \\
        Control Var. & No & No & No & No & No \\
        Weight & Region & Region  & Region & Region &Region     \\ \hline
    \end{tabular}
    }
    \begin{tablenotes}
Standard errors clustered at the country level are in parentheses. ***p \textless 0.01, **p \textless 0.05, *p \textless 0.10
\end{tablenotes}
\end{table}

Table \ref{tbl:altlag} presents the regression results with different lags of the temperature and precipitation variables. Columns (1) to (3) show the annual panel regression results, whereas columns (5) to (6) report the long-difference regression results. Columns (1) and (4) are our main specifications, which use $\mathbf{\Delta T \cdot T}$ to capture the level effects. In contrast, columns (2) and (5) use $\mathbf{\Delta T \cdot L.T}$ \textemdash the lagged temperature and precipitation variables \textemdash to capture the level effects. Columns (3) and (6) includes both contemporaneous and lagged terms by constructing an interaction variable $\mathbf{\Delta T(T+ L.T)}$. 

The results show that considering different forms of lags does not significantly affect the estimates. Specifically, columns (1) to (6) show that short-term and long-term precipitation changes consistently have no significant impact on either output level or growth. Similarly, the effects of short-term and long-term temperature changes on output levels are also limited across specifications. In contrast, the effects of both short-term and long-term temperature changes on output growth remains consistently significant across different specifications, with similar coefficient values. These results suggest that the results based on our parsimonious specifications, which include only the contemporaneous variables, are robust.

\begin{table}[!ht]
\QTR{caption}{Regression results with lags}
\label{tbl:altlag}
    \centering
    \resizebox{\textwidth}{!}{
    \begin{tabular}{>{\raggedright}p{2.5cm} c c c c c c}
    \hline
        ~ & (1) & (2) & (3) & (4) & (5) & (6) \\ 
        Model &  \multicolumn{3}{c}{Panel model} &  \multicolumn{3}{c}{Long-differece model} \\ \hline
         $\Delta T$  & -0.00588 & -0.00661 & -0.00630 & -0.351* & -0.337* & -0.344*  \\ 
        ~ & (0.0046) & (0.0047) & (0.0047) & (0.1839) & (0.1791) & (0.1816)  \\ 
        $\Delta T \cdot T$ & 0.000507 & ~ & ~ & 0.0146* & ~ & ~  \\
        ~ & (0.0003) & ~ & ~ & (0.0084) & ~ & ~  \\ 
        $\Delta T \cdot L.T$ & ~ & 0.000573 & ~ & ~ & 0.0143* & ~  \\ 
        ~ & ~ & (0.0004) &~ & ~ & (0.0083) & ~  \\
        $\Delta T(T+ L.T)$ & ~ & ~ & 0.000272 & ~ & ~ & 0.00723*  \\ 
        ~ & ~ & ~ & (0.0017) & ~ & ~ & (0.0042)  \\
        $T$ & 0.0226** & 0.0232** & 0.0229** & 0.534** & 0.533** & 0.533**  \\ 
        ~ & (0.0094) & (0.0096) & (0.0095) & (0.2355) & (0.2357) & (0.2356)  \\ 
        $T^2$ & -0.000774*** & -0.000801*** & -0.000790*** & -0.0149*** & -0.0149*** & -0.0149***  \\ 
        ~ & (0.0003) & (0.0003) & (0.0003) & (0.0055) & (0.0055) & (0.0055)  \\ 
         $\Delta P$ & -0.00288 & -0.000837 & -0.00100 & 0.105 & 0.114 & 0.107  \\ 
        ~ & (0.0094) & (0.0102) & (0.0112) & (0.4333) & (0.4335) & (0.4364)  \\ 
        $\Delta P \cdot P$ & -0.000998 & ~ & ~ & 0.0955 & ~ & ~  \\ 
        ~ & (0.0039) & ~ &~ & (0.2597) & ~ & ~  \\ 
        $\Delta P \cdot L.P$ & ~ & -0.00202 & ~ & ~ & 0.0987 & ~  \\ 
        ~ & ~ & (0.0043) & ~ & ~ & (0.2967) & ~  \\ 
        $\Delta P(P+ L.P)$ & ~ & ~ & -0.000976 & ~ &  & 0.0490  \\ 
        ~ & ~ & ~ & (0.0024) & ~ & ~& (0.1395)  \\
        $P$ & 0.0169 & 0.0153 & 0.0152 & 0.279 & 0.272 & 0.278  \\ 
        ~ & (0.0145) & (0.0135) & (0.0150) & (0.6268) & (0.6083) & (0.6190)  \\ 
        $P^2$ & -0.00405 & -0.00369 & -0.00364 & -0.115 & -0.116 & -0.116  \\ 
        ~ & (0.0035) & (0.0029) & (0.0035) & (0.1122) & (0.1133) & (0.1129)  \\ \hline
        Obs. & 41650 & 41650 & 41650 & 3332 & 3332 &3332  \\ 
        $R^2$ & 0.215 & 0.215 & 0.215 & 0.680 & 0.680& 0.680 \\ 
        Interval & 1 yr. & 1 yr. &1 yr.&10 yr. &10 yr. &10 yr. \\
        Region FE & Yes & Yes & Yes & Yes &Yes & Yes   \\
        Year FE & Yes & Yes & Yes & Yes  & Yes & Yes \\
        Region-specific time trends & Yes &Yes & Yes & No & No & No  \\
        Control Var. & No & No & No & No & No & No \\
        Weight & Region & Region  & Region & Region &Region & Region    \\ \hline
    \end{tabular}
    }
    \begin{tablenotes}
Standard errors clustered at the country level are in parentheses. ***p \textless 0.01, **p \textless 0.05, *p \textless 0.10
\end{tablenotes}
\end{table}

\begin{figure}[!ht]
\includegraphics[width=\textwidth]{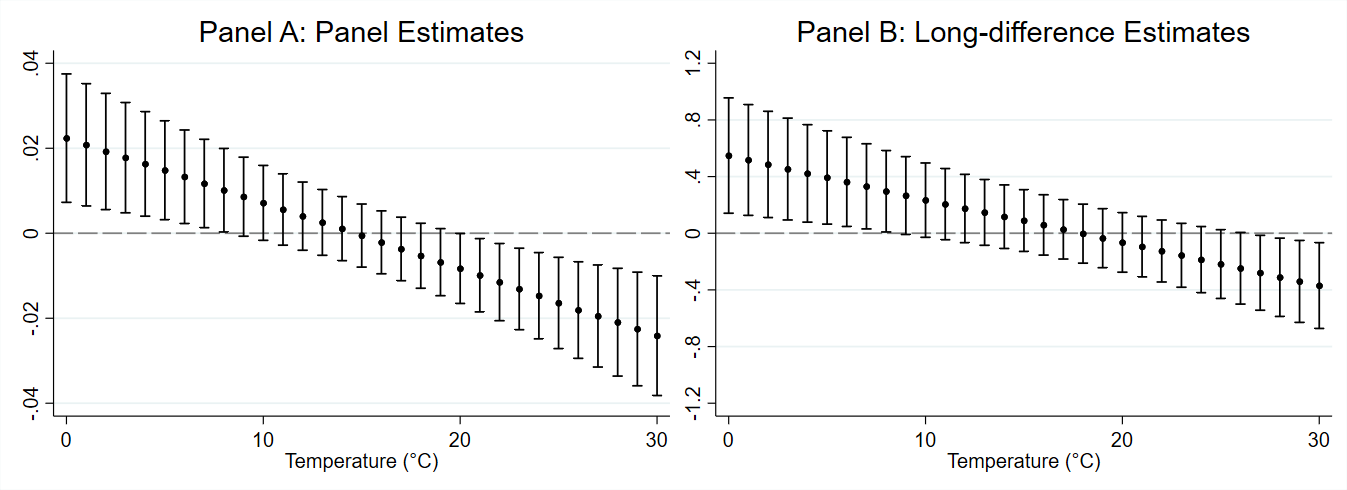}
\QTR{caption}{Bootstrap estimates results based on Panel and Long-difference models }
\label{fig:bootstrap}
\begin{figurenotes}
This figure shows the bootstrapped estimates of the marginal effects of temperature on GDP per capita growth based on the panel model (Panel A) and the long-difference model (Panel B). Dots are the median values of the 1000 bootstrapped estimates. The whiskers represent 90 percent confidence intervals.
\end{figurenotes}
\end{figure}

\textit{Bootstrap estimates.}\textemdash The bootstrap approach provides an alternative to conventional methods based on asymptotic theory, reducing reliance on distributional assumptions such as the normality of residuals or the stationary of variables \citep{horowitz2019bootstrap}. It also serves as a useful tool for Monte Carlo simulations to assess the uncertainty in the damage function. We first group our observations by country and drew 1000 samples with replacements to quantify the uncertainty of the marginal effects. Figure \ref{fig:bootstrap} presents the bootstrap estimates of the effect of temperature on output growth based on annual panel and long-difference models. Panel A, which uses the annual panel model, shows a significant marginal effect of temperature on output growth. The marginal effect of temperature at 25\celsius{} is 1.6\% and significant at 5\%. This is equal to the results from the regression in Table \ref{tbl:panel}. The long-difference estimates in Panel B also show a significant marginal effect. The marginal effect of temperature at 29\celsius{} is 34.0\% and significant at 5\%, which is just 0.9\% higher than that based on regression in Table \ref{tbl:longdiff} (33.1\% and significant 5\%). These consistent results confirm the robustness of our findings. 

\section{Discussion and Conclusion}
Quantitative estimates of climate change impact on economic output are crucial for public policy, informing decisions about investments in both emissions reductions and measures to help economies adapt to a changing climate. While existing literature has explored the relationship between climate and economic output, the findings have often been ambiguous. 

This study re-estimates the relationship between weather, climate and output through a panel long-difference model with a global subnational dataset that covers nearly all countries. We show that global coverage matters. In contrast to \citet{kalkuhl2020impact}, we find a significant effect of short-term temperature change on output growth: A 1\celsius{} increase in temperature raises annual GDP per capita growth by 1.5\% at 5\celsius{} and decreases annual GDP per capita growth by 1.6\% at 25\celsius{}. The long difference model shows a significant effect of long-term temperature change on output growth only in extremely cold and hot regions. Specifically, a 1\celsius{} increase in temperature increases annual GDP per capita growth by 3.3\% in regions with an average temperature of 5\celsius{} and decrease GDP per capita growth by 2.9\% at 29\celsius{}.

Adaptation to long-term climate change appears to be complex. When using the annual panel model, temperatures above 22\celsius{} are associated with negative impacts on economic growth. However, this threshold increases to 29\celsius{} when using the long-difference model, suggesting the presence of long-run adaptation. Meanwhile, future damage projections also highlight long-term adaptation: Annual panel estimates show that global GDP per capita is projected to decline by around 36\% by 2100 (significantly negative), while long-difference estimates show that the projected loss is only about 11\%, with the potential for positive effects. These results suggest that long-term adaptation is substantial.

However, in extremely hot regions, the story is different. According to the annual panel model, in regions with an average temperature of 29\celsius{}, a 1\celsius{} increase decreases the annual GDP per capita growth by 2.5\%. In contrast, the long-difference model estimate shows a larger annual reduction of 2.9\%. Furthermore, the bootstrap results show that in rich regions where temperatures exceed 27\celsius{}, the adaptation ratio ($1-\hat{\tau}^{LD}/\hat{\tau}^{FE}$) is significantly negative, indicating that these regions are likely to suffer more severe long-term damage from temperature increases.

Overall, our findings suggest that for majority of regions, adaptation is substantial and mitigates the negative effects of short-term temperature shocks. However, in extremely hot regions\textemdash particularly in rich, high-temperature regions\textemdash adaptation remains insufficient.

Three caveats should be considered. First, the climate is often defined as an average over 30 years \citep{auffhammer2018quantifying}. However, due to the data limitations, we could only consider ten-year averages for the long-difference regressions. We may have \emph{under}estimated adaptation. Therefore, further research is necessary to fully understand the long-term adaptation effects on climate damage mitigation when the extended time series data becomes available. Second, this study does not evaluate the potential for impacts in local regions to produce effects that affect other regions. These spillover effects may further mitigate or amplify the impacts we estimate. Therefore, further research is necessary to analyze such spillover effects of climate change on output, but this may require alternative research approaches. The presence of spatial spillover effects would imply that the residuals of the variables are neither independent nor identically distributed. Third, the literature review in the introduction reveals that people continue to propose new methods to study the impact of weather and climate, which have yet to be adopted and applied to subnational data. All that is delegated to future research.
\renewcommand{\chapter}[2]{}
\section*{REFERENCE}
\bibliographystyle{aea}
\bibliography{ref}

@ARTICLE{Tol2004,
author={Tol, R.S.J. and Verheyen, R.},
title={State responsibility and compensation for climate change damages\textemdash a legal and economic assessment},
journal={Energy Policy},
year={2004},
volume={32},
number={9},
pages={1109-1130},
doi={10.1016/S0301-4215(03)00075-2},
}

@ARTICLE{Polonik2025,
	author = {Polonik, Pascal and Ricke, Katharine and Burney, Jennifer},
	title = {Estimating the impacts of climate change: reconciling disconnects between physical climate and statistical models},
	year = {2025},
	journal = {Climatic Change},
	volume = {178},
	number = {2},
	doi = {10.1007/s10584-025-03868-w},
}

@ARTICLE{Tol2025anyas,
author={Tol, Richard S. J.},
title={Trends and biases in the social cost of carbon},
journal={Annals of the New York Academy of Sciences},
year={2025},
volume={},
number={},
pages={},
doi={},
document_type={Article},
}

@ARTICLE{Chancel2025,
	author = {Chancel, Lucas and Mohren, Cornelia and Bothe, Philipp and Semieniuk, Gregor},
	title = {Climate change and the global distribution of wealth},
	year = {2025},
	journal = {Nature Climate Change},
	volume = {15},
	number = {4},
	pages = {364 – 374},
	doi = {10.1038/s41558-025-02268-3},
}

@article{Nordhaus1993,
   author = {Nordhaus, William D.},
   title = {Rolling the `{DICE}': An Optimal Transition Path for Controlling Greenhouse Gases},
   journal = {Resource and Energy Economics},
   volume = {15},
   number = {1},
   pages = {27-50},
   year = {1993},
   type = {Journal Article}
}

@article{tol2024meta,
  title={A meta-analysis of the total economic impact of climate change},
  author={Tol, Richard S. J. },
  journal={Energy Policy},
  volume={185},
  pages={113922},
  year={2024},
  publisher={Elsevier}
}

@article{damania2020does,
  title={Does rainfall matter for economic growth? Evidence from global sub-national data (1990--2014)},
  author={Damania, Richard and Desbureaux, Sebastien and Zaveri, Esha},
  journal={Journal of Environmental Economics and Management},
  volume={102},
  pages={102335},
  year={2020},
  publisher={Elsevier}
}

@article{kalkuhl2020impact,
  title={The impact of climate conditions on economic production. Evidence from a global panel of regions},
  author={Kalkuhl, Matthias and Wenz, Leonie},
  journal={Journal of Environmental Economics and Management},
  volume={103},
  pages={102360},
  year={2020},
  publisher={Elsevier}
}

@article{burke2016adaptation,
  title={Adaptation to climate change: Evidence from US agriculture},
  author={Burke, Marshall and Emerick, Kyle},
  journal={American Economic Journal: Economic Policy},
  volume={8},
  number={3},
  pages={106--140},
  year={2016},
  publisher={American Economic Association 2014 Broadway, Suite 305, Nashville, TN 37203-2425}
}

@article{burke2015climate,
  title={Climate and conflict},
  author={Burke, Marshall and Hsiang, Solomon M and Miguel, Edward},
  journal={Annual Review of Economics},
  volume={7},
  number={1},
  pages={577--617},
  year={2015},
}

@ARTICLE{Kotz2024,
	author = {Kotz, Maximilian and Levermann, Anders and Wenz, Leonie},
	title = {The economic commitment of climate change},
	year = {2024},
	journal = {Nature},
	volume = {628},
	number = {8008},
	pages = {551 – 557},
	doi = {10.1038/s41586-024-07219-0},
}

@article{Auffhammer2022,
title = {Climate Adaptive Response Estimation: Short and long run impacts of climate change on residential electricity and natural gas consumption},
journal = {Journal of Environmental Economics and Management},
volume = {114},
pages = {102669},
year = {2022},
url = {https://www.sciencedirect.com/science/article/pii/S0095069622000432},
author = {Maximilian Auffhammer},
}

@techreport{Deryugina2017,
 title = "The Marginal Product of Climate",
 author = "Tatyana Deryugina and Solomon Hsiang",
 institution = "National Bureau of Economic Research",
 type = "Working Paper",
 series = "Working Paper Series",
 number = "24072",
 year = "2017",
 doi = {10.3386/w24072},
 URL = "http://www.nber.org/papers/w24072",
}

@article{Dell2009,
Author = {Dell, Melissa and Jones, Benjamin F. and Olken, Benjamin A.},
Title = {Temperature and Income: Reconciling New Cross-Sectional and Panel Estimates},
Journal = {American Economic Review},
Volume = {99},
Number = {2},
Year = {2009},
Month = {May},
Pages = {198-204},
URL = {https://www.aeaweb.org/articles?id=10.1257/aer.99.2.198}
}

@article{meierrieks2024temperature,
  title={Is temperature adversely related to economic development? Evidence on the short-run and the long-run links from sub-national data},
  author={Meierrieks, Daniel and Stadelmann, David},
  journal={Energy Economics},
  volume={136},
  pages={107758},
  year={2024},
  publisher={Elsevier}
}

@article{Ahmadi2025,
title = {Climate shocks, economic activity and cross-country spillovers: Evidence from a new global model},
journal = {Economic Modelling},
volume = {148},
pages = {107082},
year = {2025},
issn = {0264-9993},
doi = {https://doi.org/10.1016/j.econmod.2025.107082},
url = {https://www.sciencedirect.com/science/article/pii/S026499932500077X},
author = {Maryam Ahmadi and Chiara Casoli and Matteo Manera and Daniele Valenti},
}

@article{Neal2025,
doi = {10.1088/1748-9326/adbd58},
url = {https://dx.doi.org/10.1088/1748-9326/adbd58},
year = {2025},
month = {mar},
publisher = {IOP Publishing},
volume = {20},
number = {4},
pages = {044029},
author = {Neal, Timothy and Newell, Ben R and Pitman, Andy},
title = {Reconsidering the macroeconomic damage of severe warming},
journal = {Environmental Research Letters},
}

@article{Callahan2025,
title = {Carbon majors and the scientific case for climate liability},
journal = {Nature},
volume = {640},
pages = {893–901},
year = {2025},
url = {https://www.nature.com/articles/s41586-025-08751-3},
author = {Christopher W. Callahan and Justin S. Mankin},
}

@article{Tolfc,
  author={Richard S.J. Tol},
  title={The economic impact of weather and climate},
  journal = {Advances in Econometrics},
   volume = {},
   pages = {},
   year = {forthcoming},
   type = {Journal Article}
}

@TechReport{Bilal2024,
  author={Adrien Bilal and Diego R. K\"{a}nzig},
  title={The Macroeconomic Impact of Climate Change: Global vs. Local Temperature},
  year=2024,
  month=May,
  institution={National Bureau of Economic Research, Inc},
  type={NBER Working Papers},
  url={https://ideas.repec.org/p/nbr/nberwo/32450.html},
  number={32450},
}

@TechReport{Muller2025,
  author={Karsten M\"{u}ller and Chenzi Xu and Mohamed Lehbib and Ziliang Chen},
  title={{The Global Macro Database: A New International Macroeconomic Dataset}},
  year=2025,
  month=Apr,
  institution={National Bureau of Economic Research, Inc},
  type={NBER Working Papers},
  url={https://ideas.repec.org/p/nbr/nberwo/33714.html},
  number={33714},
}

@Article{Apergis2025,
  author={Nicholas Apergis and Mobeen Ur Rehman},
  title={{The asymmetric role of temperature deviations in economic growth: Fresh evidence from global countries and panel quantile estimates}},
  journal={International Journal of Finance \& Economics},
  year=2025,
  volume={30},
  number={1},
  pages={893-903},
  month={January},
  doi={10.1002/ijfe.2952},
  url={https://ideas.repec.org/a/wly/ijfiec/v30y2025i1p893-903.html}
}

@ARTICLE{Kikstra2021,
author={Kikstra, Jarmo S. and Waidelich, Paul and Rising, James and Yumashev, Dmitry and Hope, Chris W. and Brierley, Chris M.},
title={The social cost of carbon dioxide under climate-economy feedbacks and temperature variability},
journal={Environmental Research Letters},
year={2021},
volume={16},
number={9},
doi={10.1088/1748-9326/ac1d0b},
art_number={094037},
document_type={Article},
}

@Article{Acevedo2020,
  author       = {Acevedo, Sebastian and Mrkaic, Mico and Novta, Natalija and Pugacheva, Evgenia and Topalova, Petia},
  title        = {{The Effects of Weather Shocks on Economic Activity: What are the Channels of Impact?}},
  doi          = {10.1016/j.jmacro.2020.103},
  number       = {103207},
  volume       = {65},
  journal      = {Journal of Macroeconomics},
  year         = {2020},
}

@Article{Henseler2019,
  author={Martin Henseler and Ingmar Schumacher},
  title={The impact of weather on economic growth and its production factors},
  journal={Climatic Change},
  year=2019,
  volume={154},
  number={3},
  pages={417-433},
  doi={10.1007/s10584-019-02441-},
}

@Article{Pretis2018,
  author       = {Felix Pretis and Moritz Schwarz and Kevin Tang and Karsten Haustein and Myles R. Allen},
  title        = {Uncertain impacts on economic growth when stabilizing global temperatures at 1.5\celsius{} or 2\celsius{} warming},
  doi          = {10.1098/rsta.2016.0460},
  number       = {2119},
  pages        = {20160460},
  volume       = {376},
  journal      = {Philosophical Transactions of the Royal Society A: Mathematical, Physical and Engineering Sciences},
  year         = {2018},
}

@article{dell2012temperature,
  title={Temperature shocks and economic growth: Evidence from the last half century},
  author={Dell, Melissa and Jones, Benjamin F and Olken, Benjamin A},
  journal={American Economic Journal: Macroeconomics},
  volume={4},
  number={3},
  pages={66--95},
  year={2012},
  publisher={American Economic Association}
}

@article{burke2015global,
  title={Global non-linear effect of temperature on economic production},
  author={Burke, Marshall and Hsiang, Solomon M and Miguel, Edward},
  journal={Nature},
  volume={527},
  number={7577},
  pages={235--239},
  year={2015},
  publisher={Nature Publishing Group}
}

@article{kummu2018gridded,
  title={Gridded global datasets for gross domestic product and Human Development Index over 1990--2015},
  author={Kummu, Matti and Taka, Maija and Guillaume, Joseph HA},
  journal={Scientific data},
  volume={5},
  number={1},
  pages={1--15},
  year={2018},
  publisher={Nature Publishing Group}
}

@article{mendelsohn2006distributional,
  title={The distributional impact of climate change on rich and poor countries},
  author={Mendelsohn, Robert and Dinar, Ariel and Williams, Larry},
  journal={Environment and development economics},
  volume={11},
  number={2},
  pages={159--178},
  year={2006},
  publisher={Cambridge University Press}
}

@article{burke2018large,
  title={Large potential reduction in economic damages under UN mitigation targets},
  author={Burke, Marshall and Davis, W Matthew and Diffenbaugh, Noah S},
  journal={Nature},
  volume={557},
  number={7706},
  pages={549--553},
  year={2018},
  publisher={Nature Publishing Group UK London}
}

@article{li2022spatiotemporal,
  title={Spatiotemporal dynamics of global population and heat exposure (2020--2100): Based on improved SSP-consistent population projections},
  author={Li, Mengya and Zhou, Bing-Bing and Gao, Minyi and Chen, Yimin and Hao, Ming and Hu, Guohua and Li, Xia},
  journal={Environmental Research Letters},
  volume={17},
  number={9},
  pages={094007},
  year={2022},
  publisher={IOP Publishing}
}

@article{wang2022global,
  title={Global gridded GDP data set consistent with the shared socioeconomic pathways},
  author={Wang, Tingting and Sun, Fubao},
  journal={Scientific data},
  volume={9},
  number={1},
  pages={221},
  year={2022},
  publisher={Nature Publishing Group UK London}
}

@article{horowitz2019bootstrap,
  title={Bootstrap methods in econometrics},
  author={Horowitz, Joel L},
  journal={Annual Review of Economics},
  volume={11},
  number={1},
  pages={193--224},
  year={2019},
  publisher={Annual Reviews}
}

@article{kotz2022effect,
  title={The effect of rainfall changes on economic production},
  author={Kotz, Maximilian and Levermann, Anders and Wenz, Leonie},
  journal={Nature},
  volume={601},
  number={7892},
  pages={223--227},
  year={2022},
  publisher={Nature Publishing Group}
}

@article{kotz2021day,
  title={Day-to-day temperature variability reduces economic growth},
  author={Kotz, Maximilian and Wenz, Leonie and Stechemesser, Annika and Kalkuhl, Matthias and Levermann, Anders},
  journal={Nature Climate Change},
  volume={11},
  number={4},
  pages={319--325},
  year={2021},
  publisher={Nature Publishing Group UK London}
}

@article{song2023effects,
  title={Effects of rising and extreme temperatures on production factor efficiency: Evidence from China's cities},
  author={Song, Malin and Wang, Jianlin and Zhao, Jiajia},
  journal={International Journal of Production Economics},
  volume={260},
  pages={108847},
  year={2023},
  publisher={Elsevier}
}

@article{malpede2024long,
  title={The long-term economic effects of aridification},
  author={Malpede, Maurizio and Percoco, Marco},
  journal={Ecological Economics},
  volume={217},
  pages={108079},
  year={2024},
  publisher={Elsevier}
}

@article{gennaioli2013human,
  title={Human capital and regional development},
  author={Gennaioli, Nicola and La Porta, Rafael and Lopez-de-Silanes, Florencio and Shleifer, Andrei},
  journal={The Quarterly journal of economics},
  volume={128},
  number={1},
  pages={105--164},
  year={2013},
  publisher={MIT Press}
}

@article{klein2017anthropogenic,
  title={Anthropogenic land use estimates for the Holocene--HYDE 3.2},
  author={Klein Goldewijk, Kees and Beusen, Arthur and Doelman, Jonathan and Stehfest, Elke},
  journal={Earth System Science Data},
  volume={9},
  number={2},
  pages={927--953},
  year={2017},
  publisher={Copernicus Publications G{\"o}ttingen, Germany}
}

@article{newell2021gdp,
  title={The GDP-temperature relationship: implications for climate change damages},
  author={Newell, Richard G and Prest, Brian C and Sexton, Steven E},
  journal={Journal of Environmental Economics and Management},
  volume={108},
  pages={102445},
  year={2021},
  publisher={Elsevier}
}

@article{auffhammer2018quantifying,
  title={Quantifying economic damages from climate change},
  author={Auffhammer, Maximilian},
  journal={Journal of Economic Perspectives},
  volume={32},
  number={4},
  pages={33--52},
  year={2018},
  publisher={American Economic Association 2014 Broadway, Suite 305, Nashville, TN 37203-2418}
}

@article{tol2011social,
  title={The social cost of carbon},
  author={Tol, Richard SJ},
  journal={Annu. Rev. Resour. Econ.},
  volume={3},
  number={1},
  pages={419--443},
  year={2011},
  publisher={Annual Reviews}
}

@article{anthoff2019inequality,
  title={Inequality and the social cost of carbon},
  author={Anthoff, David and Emmerling, Johannes},
  journal={Journal of the Association of Environmental and Resource Economists},
  volume={6},
  number={2},
  pages={243--273},
  year={2019},
  publisher={University of Chicago Press Chicago, IL}
}

@article{jongman2014increasing,
  title={Increasing stress on disaster-risk finance due to large floods},
  author={Jongman, Brenden and Hochrainer-Stigler, Stefan and Feyen, Luc and Aerts, Jeroen CJH and Mechler, Reinhard and Botzen, WJ Wouter and Bouwer, Laurens M and Pflug, Georg and Rojas, Rodrigo and Ward, Philip J},
  journal={Nature Climate Change},
  volume={4},
  number={4},
  pages={264--268},
  year={2014},
  publisher={Nature Publishing Group UK London}
}

@techreport{burke2023quantifying,
  title={Quantifying climate change loss and damage consistent with a social cost of greenhouse gases},
  author={Burke, Marshall and Zahid, Mustafa and Diffenbaugh, Noah and Hsiang, Solomon M},
  year={2023},
  institution={National Bureau of Economic Research}
}

@article{kahn2021long,
  title={Long-term macroeconomic effects of climate change: A cross-country analysis},
  author={Kahn, Matthew E and Mohaddes, Kamiar and Ng, Ryan NC and Pesaran, M Hashem and Raissi, Mehdi and Yang, Jui-Chung},
  journal={Energy Economics},
  volume={104},
  pages={105624},
  year={2021},
  publisher={Elsevier}
}

@article{fankhauser2005climate,
  title={On climate change and economic growth},
  author={Fankhauser, Samuel and Tol, Richard SJ},
  journal={Resource and Energy Economics},
  volume={27},
  number={1},
  pages={1--17},
  year={2005},
  publisher={Elsevier}
}

@Article{Kolstad2020,
  author  = {Kolstad, Charles D. and Moore, Frances C.},
  title   = {Estimating the Economic Impacts of Climate Change Using Weather Observations},
  number  = {1},
  pages   = {1-24},
  url     = {https://doi.org/10.1093/reep/rez024},
  volume  = {14},
  journal = {Review of Environmental Economics and Policy},
  year    = {2020},
}

@article{Dell2014,
   author = {Dell, Melissa and Jones, Benjamin F. and Olken, Benjamin A.},
   title = {What do we learn from the weather? The new climate-economy literature},
   journal = {Journal of Economic Literature},
   volume = {52},
   number = {3},
   pages = {740-798},
   year = {2014},
   type = {Journal Article}
}

@article{letta2019weather,
  title={Weather, climate and total factor productivity},
  author={Letta, Marco and Tol, Richard S. J.},
  journal={Environmental and Resource Economics},
  volume={73},
  number={1},
  pages={283--305},
  year={2019},
  publisher={Springer}
}

@article{hsiang2016climate,
  title={Climate econometrics},
  author={Hsiang, Solomon},
  journal={Annual Review of Resource Economics},
  volume={8},
  number={1},
  pages={43--75},
  year={2016},
  publisher={Annual Reviews}
}

@article{harris2020version,
  title={Version 4 of the CRU TS monthly high-resolution gridded multivariate climate dataset},
  author={Harris, Ian and Osborn, Timothy J and Jones, Phil and Lister, David},
  journal={Scientific data},
  volume={7},
  number={1},
  pages={109},
  year={2020},
  publisher={Nature Publishing Group UK London}
}

@article{merel2021climate,
  title={Climate econometrics: Can the panel approach account for long-run adaptation?},
  author={M{\'e}rel, Pierre and Gammans, Matthew},
  journal={American Journal of Agricultural Economics},
  volume={103},
  number={4},
  pages={1207--1238},
  year={2021},
  publisher={Wiley Online Library}
}

\newpage
\appendix

\section{Mathematical Appendix: The Effects of Climate Conditions \\ Jinchi Dong, Richard S.J. Tol, Jinnan Wang}

\subsubsection{Appendix I: Pre- and Post-estimation Tests}

We first conduct the unit root test to check the stationary of variables. Since our panel data is a short panel with large cross-sections but short periods, we employ the Harris-Tzavalis test for this check. Table \ref{tbl:stationary} shows the results of unit root test. For the annual panel model, temperature and precipitation variables significantly reject the null hypothesis, confirming their stationarity. Although logarithmic GDP per capita contains a unit root, its first difference ($g_{it}$) \textemdash the dependent variable in the model \textemdash is stationary. In the panel long-difference model with ten-year interval, the tests show that all the variables are stationary. $g_{it}$ could not to be tested due to insufficient observations, but it is expected to be stationary as $ln(y_{it})$ is stationary.  

\begin{table}[!ht]
\QTR{caption}{Unit root test results}
\label{tbl:stationary}
    \centering{
    \begin{tabular}{c c c c c}
    \hline
        Model &  $ ln(y_{it})$ & $g_{it}$&$ T_{i t}$ & $ P_{i t}$   \\ \hline
        Annual Panel model & 0.786 &  0.252***& -0.0037*** & -0.0074*** \\ 
        Long-difference model &0.171*** & &-0.088*** &-0.088***  \\ \hline
    \end{tabular}}
\begin{tablenotes}
We employ the Harris-Tzavalis test for this check, which includes a time trend and subtracts cross-sectional means for annual panel model and subtracts cross-sectional means for panel long-difference model. The null hypothesis of this test is that all panels contain unit roots. The value in the table represents the $\rho$ statistic results. ***p \textless 0.01
\end{tablenotes}
\end{table}

Second, we use the Lagrange Multiplier (LM) statistic, as described by Born and Breitung (2016), to conduct the serial correlation test. The long-difference regression with a ten-year interval includes only two periods, making it impossible to conduct this test. Therefore, we focus solely on the annual panel model in this section. Table \ref{tbl:serialcor} presents the results of the test. We find that, despite the inclusion of time and region-specific time trend fixed effects, serial correlation persists in the annual panel model.

To test the impact of serial correlation on the validity of our main regression model, we first introduce a two-way clustering method, which clusters by both time and individual, to get the within-panel autocorrelation and contemporaneous cross-panel correlation robust standard error. Table \ref{tbl:HAC} presents the regression results. Column (1) shows our main specification, and column (2) presents the specification clustered by time and country. Column (2) shows that the effect of temperature and its marginal effects remain significant, although their standard errors have slightly increased. We further introduce Newey–West standard errors to account for both within-panel and cross-panel autocorrelation. Columns (3) and (4) in Table \ref{tbl:HAC} present the results with different bandwidths. The results are consistent with our main findings that the effect of temperature and its marginal effects on output growth are significant. These findings suggest that, despite the presence of some degree of serial correlation within the model, its impact on the robustness of the regression results is limited.

\begin{table}[!ht]
\QTR{caption}{Serial correlation test results}
\label{tbl:serialcor}
    \centering{
    \resizebox{\textwidth}{!}{
    \begin{tabular}{c c c c c c}
    \hline
        ~ &  First order & Second order  & Fifth order & Tenth order & Thirteenth order\\ \hline
        LM test & 8.72*** & 1.87*  &-4.88*** &-2.22** &0.13 \\ \hline
    \end{tabular}}}
\begin{tablenotes}
 We use the ``xtqptest'' command in Stata to conduct these tests. The null hypothesis for this approach is that there is no serial correlation. *p \textless 0.1, **p \textless 0.05, ***p \textless 0.01
\end{tablenotes}
\end{table}

\begin{table}[!ht]
\QTR{caption}{Serial correlation analysis for Panel models}
\label{tbl:HAC}
\centering
\resizebox{\textwidth}{!}{
    \begin{tabular}{>{\raggedright}p{2.5cm}l c c c c c}
    \hline
        ~ & (1) & (2) & (3) & (4)   \\
        Dep. var. & \multicolumn{4}{c}{Annual GDP per capita growth} \\ \hline
        $\Delta T$ & -0.00588 & -0.00588 & -0.00588 & -0.00588  \\ 
        ~ & (0.0046) & (0.0062) & (0.0066) & (0.0053)  \\
        $\Delta T \cdot T$ & 0.000507 & 0.000507 & 0.000507 & 0.000507**  \\
        ~ & (0.0003) & (0.0003) & (0.0004) & (0.0002)  \\ 
        $T$ & 0.0226** & 0.0226* & 0.0226 & 0.0226  \\ 
        ~ & (0.0094) & (0.0120) & (0.0134) & (0.0136)  \\ 
        $T^2$ & -0.000774*** & -0.000774** & -0.000774** & -0.000774**  \\
        ~ & (0.0003) & (0.0003) & (0.0003) & (0.0003)  \\
        $\Delta P$ & -0.00288 & -0.00288 & -0.00288 & -0.00288  \\ 
        ~ & (0.0094) & (0.0100) & (0.0083) & (0.0050)  \\ 
        $\Delta P \cdot P$ & -0.000998 & -0.000998 & -0.000998 & -0.000998  \\
        ~ & (0.0039) & (0.0037) & (0.0033) & (0.0020)  \\
        $P$ & 0.0169 & 0.0169 & 0.0169 & 0.0169  \\ 
        ~ & (0.0145) & (0.0164) & (0.0169) & (0.0106)  \\
        $P^2$ & -0.00405 & -0.00405 & -0.00405 & -0.00405  \\ 
        ~ & (0.0035) & (0.0039) & (0.0040) & (0.0026)  \\ \hline
        Obs. & 41650 & 41650 & 41650 & 41650  \\ 
        $R^2$ & 0.215 & 0.215 & 0.215 & 0.215   \\ 
        Region FE & Yes & Yes & Yes & Yes   \\
        Year FE & Yes & Yes & Yes & Yes \\
        Region-specific time trends & Yes & Yes & Yes  & Yes   \\
        Cluster(SE) & Country & Country,year& Region,year& Region, year \\
        Bandwidth & No & No & 2 & 10 \\
        Control Var. & No & No & No & No  \\
        Weight & Region & Region  & Region & Region    \\ \hline
        ME at 25\celsius & -0.016** & -0.016**& -0.016**& -0.016***\\ 
        ..SE & 0.0065 &  0.0073 & 0.0069 & 0.0037 \\ \hline
    \end{tabular}
}
\begin{tablenotes}
Standard errors clustered at the country level are in parentheses. ME stands for marginal effects and SE for standard error. FE denotes fixed effects. ***p \textless 0.01, **p \textless 0.05, *p \textless 0.10
\end{tablenotes}
\end{table}

\clearpage

\subsubsection{Appendix II: Data comparison}
We compared the output database used by this study (hereinafter referred to as the Kummu database) with two other widely used output databases\textemdash the Kalkuhl and World Bank databases\textemdash and discussed their differences.

The Kalkuhl database is obtained by \citet{kalkuhl2020impact}. It is a subnational level database that contains over 1518 regions in 77 countries. The time series of the database is from 1900 to 2014. However, it is a highly unbalanced panel database, with over 60\%
of regions having a duration of less than 25 years. The data stem from various statistical agencies of central or federal governments. 44 out of 77 countries used GDP data, while the others used other data to measure GDP, such as Gross Value Added (GVA) or income, to measure GDP (including countries that used these output data in certain years). The values of the data are converted to USD using market exchange rates from the FRED database of the Federal Reserve Bank of St. Louis.

World Bank database used by \citet{burke2015global, dell2012temperature} and many other
studies. It is obtained from the World Bank World Development Indicators database \footnote{https://databank.worldbank.org/reports.aspx?source=World-Development-Indicators}. It is a country-level database that contains 266 countries from 1960 to 2021. The values of
the data are converted to 2017 constant international US dollars (2017 PPP). 

Table \ref{tbl:compdescrip} shows the descriptive statistics of Kummu, Kalkuhl and the World Bank database. To ensure comparability among these three databases, we aggregated the Kummu and Kalkuhl data from the subnational to the country level and focused on the shared countries and years. We find that the Kummu and World Bank databases share similar statistical results. The differences between the means of GDP per capita growth rates in the Kummu and World Bank databases are also statistically insignificant. However, due to inconsistent data sources and different currency conversion methods, the GDP per capita data in the Kalkuhl database show a significant difference from the data in the Kummu and World Bank databases.

\begin{table}[!ht]
\QTR{caption}{Descriptive statistics of three output databases}
\label{tbl:compdescrip}
    \centering
    \begin{tabular}{ccccc}
    \hline
        ~ & ~ & (1) & (2) & (3)  \\ 
        \multicolumn{2}{c}{ Database} & Kummu & Kalkuhl & World Bank  \\ \hline
        \multicolumn{2}{c}{ Mean} & 0.025 & 0.065 & 0.027  \\ 
        \multicolumn{2}{c}{ S.D.} & 0.042 & 0.132 & 0.039  \\ 
        \multicolumn{2}{c}{ Min} & -0.232 & -0.846 & -0.171  \\ 
        \multicolumn{2}{c}{ Max} & 0.336 & 1.134 & 0.236  \\ \hline
        \multicolumn{2}{c}{ Obs.} & \multicolumn{3}{c}{1103} \\
        \multicolumn{2}{c}{ Countries} & \multicolumn{3}{c}{77} \\ 
        \multicolumn{2}{c}{ Years} & \multicolumn{3}{c}{1991-2014} \\ \hline
        \multirow{3}{*}{t test} & Kummu & ~ & -9.89*** & -1.07  \\
        ~ & Kalkuhl & ~ & ~ & 9.50***  \\ 
        ~ & World Bank & ~ & ~ &   \\ \hline
    \end{tabular}
    \begin{tablenotes}
The alternative hypothesis of t test is $H_{a}:\mu \neq \mu_0$. ***p \textless 0.01, **p \textless 0.05, *p \textless 0.10.
\end{tablenotes}
\end{table}

The difference is consistent when we conducted regression and correlational analyses. We used fixed-effects panel regression and Pearson and Spearman tests to reveal the relations between these three databases. Table \ref{tbl:compreg} shows the fixed-effects panel regression results. We use the output data from each database as the dependent variable and the output data from the other two databases as independent variables in these regressions. Columns (1) to (6) show that all regression results are statistically significant. The coefficients between the Kummu and World Bank databases are all above 0.87, approaching 1, with high $R^2$ values exceeding 0.8 (columns 2 and 4). In contrast, the coefficients between the Kalkuhl database and the other two databases show significant deviations from 1, with $R^2$ values around 0.5. These results indicate substantial differences in interannual output variation trends between the Kalkuhl database and the other two datasets.

Table \ref{tbl:compcorr} presents the Pearson and Spearman correlation test results for the three databases.  Correlation coefficients approaching $\pm$1 denote stronger correlations. The results in Table \ref{tbl:compcorr} show significant correlations among all three databases, with particularly strong associations between the Kummu and World Bank databases (r \textgreater 0.85). In contrast, the Kalkuhl database demonstrates weaker correlations with both the Kummu and World Bank databases (r $\approx$ 0.5).

\begin{table}[!ht]
\QTR{caption}{Fixed-effects regression results between three output databases}
\label{tbl:compreg}
    \centering
    \resizebox{\textwidth}{!}{
    \begin{tabular}{ccccccc}
    \hline
        ~ & (1) & (2) & (3) & (4) & (5) & (6)  \\
        Dep var. & Kummu & Kummu & Kalkuhl & Kalkuhl & World Bank & World Bank \\ \hline
        Kummu & ~ & ~ & 0.886*** &~&0.743*** &~  \\ 
        ~ & ~&~&(0.106) & ~ & (0.017) &~  \\ 
        Kalkuhl & 0.0729*** &~&~&~&~& 0.094***    \\ 
        ~ &  (0.009) & ~& ~&~&~& (0.007)   \\ 
        World Bank & ~& 0.875***&~&1.304***&~&~ \\ 
        ~&~&(0.020)&~&(0.112)&~&~ \\ \hline
        Obs. & 1103 & 1103 & 1103&1103&1103&1103  \\ 
        $R^2$ & 0.49 & 0.81 & 0.41 &0.44&0.82&0.54  \\ 
        Region FE & YES & YES & YES &YES &YES &YES \\ 
        Year FE & YES & YES & YES  &YES &YRS &YES\\ \hline
    \end{tabular}}
    \begin{tablenotes}
FE denotes fixed effects. Clustered standard errors at the country level are in parentheses. ***p \textless 0.01, **p \textless 0.05, *p \textless 0.10.
\end{tablenotes}
\end{table}

\begin{table}[!ht]
\QTR{caption}{Correlation analysis results between three output databases}
\label{tbl:compcorr}
    \centering
    \resizebox{\textwidth}{!}{
    \begin{tabular}{ccccccc}
    \hline
        ~ & \multicolumn{3}{c}{ Pearson Correlation} & \multicolumn{3}{c}{ Spearman Correlation}\\
        ~ & Kummu & Kalkuhl & WorldBank & Kummu & Kalkuhl & WorldBank  \\ \hline
        Kummu & 1.00 & 0.48*** & 0.85*** & 1.00 & 0.52*** & 0.89***  \\ 
        Kalkuhl & ~ & 1.00 & 0.41*** & ~ & 1.00 & 0.49***  \\ 
        WorldBank & ~ & ~ & 1.00 & ~ & ~ & 1.00  \\ \hline
    \end{tabular}}
\end{table}

Overall, these findings suggest that the Kummu, Kalkuhl, and World Bank databases exhibit measurable correlations. However, the correlation between the Kummu and World Bank databases is substantially stronger than the Kalkuhl database. Furthermore, the World Bank data are derived from official government statistical agencies and international organizations, having undergone standardized methodological adjustments that ensure their authoritative status. Thus, the high correlation between the Kummu and World Bank databases further validates the reliability of the Kummu dataset used in this study.

Compared with the above two output per capita databases, the Kummu database used in our study has three advantages. First, given the high heterogeneity of climate and economic activities, data aggregated at the country level tend to dilute or completely mask some useful information \citep{damania2020does}. The subnational database provided by Kummu allows for a more detailed spatial description of climate and economic variables. Second, the Kummu database covers almost all countries worldwide, while Kalkuhl omitted the vast majority of countries in Africa, Southeast Asia and Central America. These omitted regions happen to be hot and less developed. Omitting these regions may cause biased estimates. Third, the Kummu database uses GDP data for all years. This homogeneous data allows for more accurate estimates.

\end{document}